\documentclass[11pt,a4paper]{article}

\usepackage{hyperref}
\usepackage{a4wide}
\usepackage{amsmath,amsthm}
\usepackage[latin1]{inputenc}
\usepackage[all]{xy}

\usepackage{amsmath,amsfonts}
\usepackage{latexsym}

\def \eref#1 {({\ref{#1}}) }

\def \cbb{\mathbb{C}}

\def \rbb{\mathbb{R}}

\def \zbb{\mathbb{Z}}

\def \bcal {\mathcal{B}}
\def \ccal {\mathcal{C}}

\def \ecal {\mathcal{E}}
\def \fcal {\mathcal{F}}
\def \gcal {\mathcal{G}}
\def \hcal {\mathcal{H}}

\def \kcal {\mathcal{K}}

\def \ocal {\mathcal{O}}
\def \pcal {\mathcal{P}}

\def \scal {\mathcal{S}}

\def \vcal {\mathcal{V}}

\def \afk  {\mathfrak{A}}
\def \bfk  {\mathfrak{B}}

\def \ffk  {\mathfrak{F}}

\def \zfk  {\mathfrak{Z}}

\def \xv{\vec x}

\def \<  {\langle}
\def \>  {\rangle}

\def \. { \,\! }

\def \cdotarg { \, \cdot \, }

\def \id {\mathrm{id}}
\def \idop {\mathbf{1}}

\def \csq {\quad \Rightarrow \quad}

\def \equivalent {\quad \Leftrightarrow \quad}

\def \st {^\ast}
\def \isom {\cong}
\def \restrict {\lceil}

\newsymbol \subsetneqq 2324
\newsymbol \subsetneq 2328

\def \supp {\mathrm{supp}\,}

\def \expltext#1 {\\ \text{\footnotesize{ (#1) }}\\}
\def \intercomm#1 {\\ \text{\footnotesize{ (#1) }}\\}
\def \undercomm#1 {\underset{\text{\scriptsize{ (#1) }}}}
\def \overcomm#1 {\overset{\text{\scriptsize{ (#1) }}}}

\newcommand{\hrskp}[2]{( #1 | #2 ) }
\newcommand{\bighrskp}[2]{\big( #1 \,\big|\, #2 \big) }

\def \boundedops {\bfk(\hcal)}

\def \dim {\mathrm{dim}\,}
\def \ad  {\mathrm{ad}\,}

\def \tracenorm#1 { \| #1 \| _1 }
\def \hsnorm#1 { \| #1 \| _2 }

\def \vectorcomp#1 {
  \left( \begin{array}{c}
  #1
  \end{array}
  \right)  }

\def \pder#1#2 { \frac{ \partial #1 }{ \partial #2 }}

\newcommand{\localitemlabels}{
  \renewcommand{\theenumi}{(\roman{enumi})}
  \renewcommand{\labelenumi}{\theenumi}
}

\newcommand{\mkraum}{ \rbb^{s+1} }

\newcommand{\poincare}{\pcal_+^{\uparrow}}

\newcommand{\normalpart}{\mathsf{N}}

\newcommand{\universal}{_\mathrm{u}}

\DeclareMathOperator{\rank}{rank}
\DeclareMathOperator{\img}{img}

\newcommand{\uomega}{\underline{\omega}}
\newcommand{\uA}{\underline{A}}

\newcommand{\uB}{\underline{B}}
\newcommand{\uC}{\underline{C}}
\newcommand{\up}{\underline{p}}
\newcommand{\uR}{\underline{R}}
\newcommand{\uafk}{\underline{\afk}}
\newcommand{\ubfk}{\underline{\bfk}}
\newcommand{\ualpha}{\underline{\alpha}}
\newcommand{\udelta}{\underline{\delta}}
\newcommand{\usigma}{\underline{\sigma}}
\newcommand{\uSigma}{\underline{\Sigma}}
\newcommand{\uphi}{\underline{\phi}}
\newcommand{\uPhi}{\underline{\Phi}}
\newcommand{\ugcal}{\smash{\underline{\gcal}}}

\newcommand{\oprod}{\underline{\Pi}}

\newcommand{\wtens}{\bar{\otimes}}

\def \xv {\mathbf{x}}

\newcommand{ \cinfty } {  \ccal^\infty }
\newcommand{ \cinftyh } {  \ccal^\infty (\hcal) }
\newcommand{ \cinftys } {  \ccal^\infty(\Sigma) }
\newcommand{ \cinftyss } {  \ccal^\infty(\Sigma)^\ast }

\newcommand{\evp} { ^\lbrack \!\,' \!\,^\rbrack }

\newcommand{\lnorm}[2]{  \| #1 \|^{( #2 )}  }

\def \PhiFH { \Phi_{ \mathrm{FH} }  }
\def \PhiFHLim { \Phi_{ \mathrm{FH},0 }  }
\def \PhiFHHat { {\hat\Phi}_{ \mathrm{FH} }  }

\newcommand{\norm}[1]{  \| #1 \|  }

\newcommand{\mkr}{\rbb^{s+1}}

\def \cdotarg { \, \cdot \, }

\def \< {\langle}
\def \> {\rangle}

\def \id {\mathrm{id}}

\def \csq {\quad \Rightarrow \quad}

\def \equivalent {\quad \Leftrightarrow \quad}

\def \st {^\ast}
\def \restrict {\lceil}

\newcommand{\mtxe}[3]{ \hrskp{#1}{#2 | #3} }

\def \boundedops {\bcal(\hcal)}

\def \dim {\mathrm{dim}\,}
\def \ad  {\mathrm{ad}\,}

\def \tracenorm#1 { \| #1 \| _1 }
\def \hsnorm#1 { \| #1 \| _2 }

\newcommand{\wstar}{weak-$\ast$ }

\newcommand{\bigsetprop}[2]{ \big\{ #1 \, \big| \, #2 \big\} }

\newcommand{\mean}{\mathsf{m}}
\newcommand{\convol}{\mathsf{K}}

\newtheorem{definition}{Definition}
\newtheorem{lemma}[definition]{Lemma}
\newtheorem{proposition}[definition]{Proposition}
\newtheorem{theorem}[definition]{Theorem}
\newtheorem{corollary}[definition]{Corollary}

\newtheorem{psc}{Phase space condition}

\numberwithin{equation}{section}
\numberwithin{definition}{section}

\newcommand{\inst}[1]{$^\textrm{#1}$ }
\newcommand{\email}[1]{e-mail: #1}

\newcommand{\cmpqed}{}

\begin{document}

\title{Scaling Algebras and Pointlike Fields: 
A Nonperturbative Approach to Renormalization\thanks{%
Work supported by MIUR, GNAMPA-INDAM, and the EU network
``Quantum Spaces -- Non Commutative Geometry'' (HPRN-CT-2002-00280).}%
}
\author{Henning Bostelmann\inst{1}
\and Claudio D'Antoni\inst{2}
\and Gerardo Morsella\inst{3}}

\date{
\parbox[t]{0.9\textwidth}{\footnotesize{%
\begin{enumerate}
\renewcommand{\theenumi}{\arabic{enumi}}
\renewcommand{\labelenumi}{\theenumi}
\item Dipartimento di Matematica, Università di Roma ``Tor Vergata'',
Via della Ricerca Scientifica, 00133 Roma, Italy,
\email{bostelma@mat.uniroma2.it}
\item Dipartimento di Matematica, Università di Roma ``Tor Vergata'',
Via della Ricerca Scientifica, 00133 Roma, Italy, \email{dantoni@mat.uniroma2.it} 
\item Scuola Normale Superiore di Pisa, Piazza dei Cavalieri, 7, 56126 Pisa, Italy, \email{gerardo.morsella@sns.it}
\end{enumerate}
}}
\\
October 8, 2008}

\maketitle

\begin{center}
\emph{Dedicated to Klaus Fredenhagen on the occasion of his 60th birthday.}
\end{center}

\begin{abstract}
We present a method of short-distance analysis in quantum field theory
that does not require choosing a renormalization prescription a priori.
We set out from a local net of algebras with associated pointlike quantum fields.
The net has a naturally defined scaling limit in the sense of Buchholz and Verch;
we investigate the effect of this limit on the pointlike fields.
Both for the fields and their operator product expansions, a well-defined limit
procedure can be established. This can always be interpreted in the usual sense
of multiplicative renormalization, where the renormalization factors are 
determined by our analysis. We also consider the limits of symmetry actions.
In particular, for suitable limit states, the group of scaling transformations
induces a dilation symmetry in the limit theory.
\end{abstract}

\section{Introduction}\label{introSec}

Renormalization has proven to be one of the key concepts of quantum field theory,
in particular in the construction of models. We can roughly divide its 
mathematical and physical implications as follows, even though
they are often mixed in one approach.

First, renormalization has a \emph{constructive} aspect: It serves as a tool to remove
divergencies, momentum-space cutoffs, or lattice restrictions from unphysical theories
in order to arrive at a physical limit theory. This aspect is found both in perturbative 
approaches and in mathematically rigorous constructions of quantum field theory, 
usually in the Euclidean regime \cite{FRS:where_we_are}.

Second, renormalization is a means of \emph{short distance analysis:} It allows to 
pass from a given physical theory to a theory which describes the behavior at 
short distances, often expected to be simpler than the full theory, 
in particular when this limit theory is a model of free particles
(asymptotic freedom). This is the idea that underlies quantum chromodynamics 
and the parton picture in high energy physics. 

It is the second aspect that we are interested in here. We assume that a fully constructed,
mathematically rigorous quantum field theory ``at finite scales'' is given, 
and investigate its short distance behavior in a physically natural setting.
As Buchholz and Verch have shown \cite{BucVer:scaling_algebras}, it is possible
to define the short-distance limit of such a theory in a model-independent way.
This method is based on the algebraic approach to quantum field theory \cite{Haa:lqp},
and allows applications in particular to the charge structure of the theory 
in the scaling limit \cite{Buc:quarks,DMV:scaling_charges,DAnMor:supersel_models}.

This algebraic approach to renormalization does not depend on technical details of the theory,
such as a choice of generating quantum fields. In fact, it is not even necessary to make any
reference to pointlike localized quantum fields at all, or even assume their existence; 
the renormalization limit can be defined
referring only to the algebras of bounded operators associated with finite regions. On the other
hand, it is not obvious how this approach relates to pointlike quantum fields, given that they
exist in the theory, and how the usual picture of point field renormalization emerges.

The present work aims at clarifying these questions. We set out from a theory given as a local net
of algebras, but assume that pointlike fields are associated to these algebras in the way proposed
in \cite{Bos:short_distance_structure}. Then we consider the scaling limit of the theory in the sense
of Buchholz and Verch, and analyze its effect on the pointlike fields. 

Specifically, the description of fields in \cite{Bos:short_distance_structure} is based on a certain
phase space condition. We show that, given that this condition holds at finite scales, 
it carries over to the scaling limit theory, so that also the limit theory has a well-described
connection with pointlike quantities. 
We analyze in detail how limits of pointlike objects arise from the algebras,
and show that a multiplicative renormalization of quantum fields naturally follows
as a consequence of our setting.

We also discuss the effect of renormalization transformations on the operator product expansion
in the sense of Wilson \cite{Wil:non-lagrangian,WilZim:products}, which is known to have a precise
meaning at finite scales \cite{Bos:product_expansions}. The operator product expansion 
plays an important role in this context: It reflects 
how renormalization transformations change the interaction of the theory, 
which are captured here in the structure constants of the ``improper algebra'' of pointlike fields.

On a heuristic level, our method can be understood as follows:
In the usual field-theoretic setting, renormalization of a pointlike field $\phi(x)$
is established by a purely geometric scaling in space-time, combined with a 
multiplication by a $c$-number $Z_\lambda$ depending on scale.
The field at scale $\lambda$ is given by
\begin{equation}
   \phi_\lambda(x) = Z_\lambda \phi(\lambda x),
\end{equation}
where e.g. for a real scalar free field in physical space-time, one would choose
$Z_\lambda = \lambda$. This $\phi_\lambda$ converges to a limit field $\phi_0$,
e.g. in the sense of Wightman functions or of suitable matrix elements.
The choice of $Z_\lambda$ is not unique, and may contain ambiguities to some extent,
even if these do not influence the structure of the limit theory for a free field.

In the algebraic setting, one considers the set of \emph{all} such possible 
renormalization schemes, without selecting a preferred choice. One abstractly works with
functions $\lambda \mapsto \uA_\lambda$, valued in the bounded operators, that 
are subject only to a geometric condition, $\uA_\lambda \in \afk(\lambda \ocal)$ for some
region $\ocal$, and to a continuity condition that serves to keep the unit of action
constant in the limit. A typical function may be thought of as
\begin{equation}
   \uA_\lambda = \exp( i Z_\lambda \phi(f_\lambda) ), \quad
 f_\lambda(x) = \lambda^{-4} f(\lambda^{-1}x),
\end{equation}
where $f$ is a fixed test function. (This is up to a necessary smearing in space-time.) 
However, the explicit form of $Z_\lambda$ needs not explicitly be fixed in this approach, 
since it does not influence the norm of $\uA_\lambda$.

The important point of our analysis is now that the $Z_\lambda$ can be constructed from 
the algebraic setting, rather than fixing them from the outset. Namely, 
following \cite{Bos:short_distance_structure}, the link between fields and bounded operators
is given as follows: Bounded operators $A$ localized in a double cone $\ocal_r$ of radius $r$
centered at the origin can be approximated by local fields by means of a series expansion,
\begin{equation}
  A \approx \sum_j \sigma_j(A) \phi_j,
\end{equation}
where $\sigma_j$ are normal functionals and $\phi_j$ are quantum fields localized at $x=0$,
both independent of $r$. The sum is to be understood as an asymptotic series in $r$. 
For example, for the real scalar free field in $3+1$ dimensions, the first terms of the expansion are
\begin{equation}
  A \approx \hrskp{\Omega}{A\Omega} \idop + \sigma_1(A) \phi,
\end{equation}
where $\phi$ is the free field and $\sigma_1$ a certain matrix element with 1-particle functions
(cf. \cite[Eq.~A19]{Bos:short_distance_structure}). Now this $\sigma_1$ has the property
that $\|\sigma_1 \restrict \afk(\ocal_r)\| \sim r$. Thus inserting operators $\uA_\lambda
\in \afk(\lambda \ocal)$ with some fixed region $\ocal$, we obtain
$\sigma_1(\uA_\lambda) \sim \lambda$, and can expect that $Z_\lambda := \sigma_1(\uA_\lambda)$
is a suitable renormalization factor for the field $\phi$. 
We shall see in Sec.~\ref{fieldSec} that this heuristic expectation
can indeed be made precise. The central point here is that the factors $Z_\lambda$ arise
as a \emph{consequence} of our analysis; they are determined by the scaling limit of the algebras,
it is not necessary to put them in explicitly.

Another important aspect of our analysis is the description of symmetries, in particular dilations.
It is heuristically expected that scaling transformations, which would map $\uA_\lambda$ to
$\uA_{\mu\lambda}$, or $Z_\lambda \phi$ to $Z_{\mu\lambda} \phi$, relate to a dilation 
symmetry in the limit theory. In order to make this precise, we need to generalize 
the structures introduced in \cite{BucVer:scaling_algebras}, since the limit states 
considered there are not invariant under scaling. We propose more general limit states
that are in fact invariant under scaling transformations of the above kind, and allow
a canonical implementation of these transformations in the limit, yielding 
an action of the dilation group. However, these generalized limit states are no longer pure
states; and pure limit states are not dilation covariant.

The paper is organized as follows: In Sec.~\ref{scalingSec}, we recall the algebraic
approach to renormalization, and introduce the generalizations needed for our analysis.
This includes the dilation invariant limit states mentioned above, but also a 
generalization of the scaling limit from bounded operators to unbounded objects.
Section~\ref{fieldSec} then describes pointlike fields associated with the theory,
and analyzes their scaling limit, giving a construction for the renormalization factors $Z_\lambda$.
Section~\ref{opeSec} concerns operator product expansions and their scaling limits.
We end with a conclusion in Sec.~\ref{conclusionsSec}, in particular discussing the
expected situation in quantum chromodynamics. In the appendix, we handle a technical construction
regarding states on $C\st$ algebras.

One notational convention applies throughout the paper: In order to avoid complicated index notation,
we will sometimes write the index of a symbol in brackets following it;
e.g. we write $\alpha[x,\Lambda]$ as a synonym for $\alpha_{x,\Lambda}$.

\section{The algebraic approach to renormalization} \label{scalingSec}

We use the algebraic approach to quantum field theory \cite{Haa:lqp} for our analysis.
Let us briefly summarize the basic structure, since we will use several variants of it:
A \emph{net of algebras} is a map $\afk: \ocal \mapsto \afk(\ocal)$ 
that assigns to each open bounded subset $\ocal \subset \mkraum$ of Minkowski space
a $C^*$ algebra $\afk(\ocal)$, such that \emph{isotony} holds, i.e.
$\afk(\ocal_1) \subset \afk(\ocal_2)$ if $\ocal_1\subset \ocal_2$.
For such a net, we can define the \emph{quasilocal algebra}, again denoted by $\afk$ following usual convention,
as the closure (or inductive limit) of $\cup_\ocal \afk(\ocal)$, where the union
runs over all open bounded regions. 

\newcounter{algdef}

\begin{definition} \label{netDef}
Let $\gcal$ be a Lie group of point transformations of Minkowski space that includes the 
translation group. A \emph{local net of algebras} with symmetry group $\gcal$
is a net of algebras $\afk$ together with a representation $g \mapsto \alpha_g$ of $\gcal$ 
as automorphisms of $\afk$, such that
\begin{enumerate}
\localitemlabels
\item $[A_1, A_2] = 0$ if $\ocal_1,\ocal_2$ are two spacelike separated regions,
and $A_i \in \afk(\ocal_i)$ \emph{(locality)};
\item  $\alpha_{g} \afk(\ocal) = \afk(g.\ocal)$
for all $\ocal,g$ \emph{(covariance)}.
\end{enumerate}
We call $\afk$ a \emph{net in a positive energy representation} if, in addition, the $\afk(\ocal)$
are $W^*$ algebras acting on a common Hilbert space $\hcal$, and
\begin{enumerate}
\setcounter{enumi}{2}
\localitemlabels
\item there is a strongly continuous unitary representation $g \mapsto U(g)$ of $\gcal$
on $\hcal$ such that $\alpha_{g} = \ad U(g)$;
\item the joint spectrum of the generators of translations $U(x)$ lies
in the closed forward light cone $\bar \vcal_+$ \emph{(spectrum condition)};
\item there exists a vector $\Omega \in \hcal$ which is invariant under all $U(g)$ and
  cyclic for $\afk$.
\end{enumerate}
We call $\afk$ a \emph{net in the vacuum sector} if, in addition, 
\begin{enumerate}
\setcounter{enumi}{5}
\localitemlabels
\item \label{uniqVacCond} the vector $\Omega$ is unique (up to scalar factors) as an invariant vector for the translation group.
\end{enumerate}
\end{definition}

We are frequently interested in special regions $\ocal$, namely standard double cones $\ocal_r$
of radius $r$ centered at the origin. For their associated algebra $\afk(\ocal_r)$, 
we often use the shorthand notation $\afk(r)$.

The group $\gcal$ will usually be the (proper orthochronous) Poincaré group $\poincare$,
but in some cases additionally include the dilations. In a slight abuse of notation, we will sometimes
refer to translations as $\alpha_x$ or $U(x)$, to Lorentz transforms as $\alpha_\Lambda$ or $U(\Lambda)$,
etc., leaving out those components of the group element that equal the identity of the corresponding subgroups.

In the Hilbert space case, we write the positive generator of time translations as $H$, and its spectral projectors as $P(E)$. We denote the vacuum state as $\omega = \mtxe{\Omega}{\cdotarg}{\Omega}$.
It the case of a vacuum sector, it follows from condition~\ref{uniqVacCond} that $\omega$ is a pure state.

\subsection{Scaling algebra}

Our approach to renormalization in the context of algebraic quantum field theory is
based on the results of Buchholz and Verch \cite{BucVer:scaling_algebras}, however with some
modifications. Let us briefly recall
the notions introduced there.

We assume in the following that a theory ``at scale 1'', denoted by $\afk$, is given,
and fulfills the requirements of Definition~\ref{netDef} for a local net in the vacuum sector,
with symmetry group $\poincare$. 
We will analyze the scaling limit of this theory.
To this end, we define the set of ``scaling functions'',
\begin{equation}
  \ubfk := \{ \uB : \rbb_+ \to \boundedops \,|\, \sup_\lambda \|\uB_\lambda \| < \infty \},
\end{equation}
where we write $\uB_\lambda$ rather than $\uB(\lambda)$ for the image points. 
Equipping $\ubfk$ with pointwise addition, multiplication,
and $\ast$ operation, and with the norm $\|\uB\| = \sup_\lambda \|\uB_\lambda\|$, it is easily
seen that $\ubfk$ is a $C\st$ algebra.

On $\ubfk$, we introduce an automorphic action $\ualpha$ of the Poincaré group
$\poincare$ by
\begin{equation}
    (\ualpha_{x,\Lambda}(\uB))_\lambda := \alpha_{\lambda x, \Lambda}(\uB_\lambda).
\end{equation}
We also have an automorphic action $\udelta$ of the dilation group $\rbb_+$ on
$\ubfk$:
\begin{equation}
    (\udelta_{\mu}(\uB))_\lambda := \uB_{\mu\lambda}.
\end{equation}
It is easily checked that the $\ualpha_{x,\Lambda}$ and $\udelta_\mu$ fulfill the usual commutation rules.
We will combine them into a larger Lie group $\ugcal$, with elements
$g=(\mu,x,\Lambda)$, and their representation denoted by $\ualpha$ again:
\begin{equation}
    \ualpha_g = \ualpha_{\mu,x,\Lambda} := \udelta_\mu \circ \ualpha_{x, \Lambda} 
 =  \ualpha_{\mu x, \Lambda} \circ \udelta_\mu.
\end{equation}

We are now ready to define a new set of local algebras as subalgebras of $\ubfk$:
\begin{equation}
  \uafk(\ocal) := \big\{ \uA \in \ubfk \,|\, \uA_\lambda \in \afk(\lambda \ocal) \text{ for all }\lambda>0; \; 
   g \mapsto \ualpha_g( \uA ) \text{ is norm continuous} \big\}.
\end{equation}
This defines a new local net of algebras in the sense of Definition~\ref{netDef},
with symmetry group $\ugcal$,
referring to its usual geometric action. 
We denote by $\uafk$ the associated quasilocal algebra,
i.e.~the norm closure of $\cup_\ocal \uafk(\ocal)$; 
our interest is actually in its Hilbert space representations. 
Note that $\uafk$ has a large center $\zfk(\uafk)$, consisting of those $\uA\in\uafk$
where each $\uA_\lambda$ is a multiple of $\idop$. This will turn out to be important
in our analysis.

Let us recall from \cite{BucVer:scaling_algebras} what the two conditions on $\uA\in\uafk(\ocal)$
heuristically stand for: $\uA_\lambda \in \afk(\lambda \ocal)$ means that our scaled operators
are localized in smaller and smaller regions, as required for the scaling limit. The norm continuity 
of $g \mapsto \ualpha_g( \uA )$ ensures that the unit of action $\hbar$ is kept constant in this limit.
We are requiring a bit more here than in \cite{BucVer:scaling_algebras}, inasmuch as also the action
of dilations is required to be norm continuous; so we are actually considering subalgebras of those
investigated by Buchholz and Verch. This will not influence the construction, but allow us to implement
a continuous action of the dilation group in the limit theory later.
We note that the continuity conditions imposed are not too strong restrictions, since operators
$\uA$ fulfilling them can be constructed in abundance. In order to show this, we need some technical preparations, which 
will be useful also in the following. 

We denote here by $\bcal(\rbb_+)$ and $\ccal_b(\rbb_+)$ the $C^*$-algebras of bounded functions 
and of bounded continuous functions on $\rbb_+$ respectively, equipped with the supremum norm $\|f\|_\infty = \sup_{\lambda>0}|f(\lambda)|$.

\begin{lemma}\label{lem:meanconvolution}
Let $a>1$, and let $f\in \ccal_b(\rbb_+)$ with $\supp f \subset (1/a,a)$. There exists a bounded linear operator 
$\convol_f: \bcal(\rbb_+)
\to \bcal(\rbb_+)$ such that:
\begin{enumerate}
\localitemlabels
\item  \label{convolNorm} 
$| (\convol_f g)(\lambda)| \leq 2 \log a\; \|f\|_\infty \sup_{\mu\in[1/a,a]}|g(\lambda\mu)|$
for all $\lambda > 0$; in particular, $\|\convol_f\| \leq 2 \|f\|_\infty \log a$ ;
\item  \label{convolLin} 
at fixed $a$, the map $f \mapsto \convol_f$ is linear;
\item \label{convolShift}
if $\mu >0$ is such that $\mu^{-1} \,\supp f \subset (1/a,a)$, then 
 $\convol[f(\mu \cdotarg)]g(\mu\cdotarg) = \convol[f] g$;
\item \label{convolCont} 
for each $g\in \bcal(\rbb_+)$, the map $\mu \in \rbb_+ \mapsto (\convol_f g)(\mu\,\cdot) \in \bcal(\rbb_+)$ is
continuous;
\item \label{convolIntegral}
if $g \in \ccal_b(\rbb_+)$, then $(\convol_f g)(\lambda) = \int_{\rbb_+} f(\mu)g(\lambda\mu) \, d\mu/\mu$.
\end{enumerate}
\end{lemma}

\begin{proof}
Using the map $\lambda \in [1/a,a]\mapsto \log_a \lambda \in [-1,1]$, and considering $[-1,1]$
with endpoints identified,  we can endow the interval $[1/a,a]$ with the structure of 
an abelian group under multiplication. There 
exists therefore an invariant mean $\mean_a$ over the Banach space of bounded functions
on $[1/a,a]$, i.e.\ a bounded linear functional on this space such that $\mean_a(\idop)=1$, $\mean_a(h) \geq 0$ for $h \geq 0$, 
and $\mean_a(h(\mu\,\cdot)) = \mean_a(h)$ for all $\mu \in [1/a,a]$. We define then
\begin{equation}
(\convol_f g)(\lambda) := 2 \mean_a\big(f(\cdot)g(\lambda\,\cdot)\big) \log a .
\end{equation}
Properties \ref{convolNorm}, \ref{convolLin}, and  \ref{convolShift} then
follow from boundedness, linearity, and invariance of the mean $\mean_a$, respectively.
For property \ref{convolCont}, we first note that
it is sufficient to show continuity at $\mu=1$. For $\mu$ close enough to 1,
we can use \ref{convolNorm}--\ref{convolShift} to show
\begin{equation}\label{eq:dilationcont}
\| (\convol_f g)(\mu\,\cdot)- \convol_f g\| \leq 2\log a\,\|f(\mu^{-1}\,\cdot)-f\|_\infty \|g\|_\infty.
\end{equation}
The right hand side converges to 0 as $\mu \to 1$, thanks to the uniform continuity of $f$,
which proves \ref{convolCont}.
Finally, by uniqueness
of the (normalized) Haar measure on $[-1,1]$ we immediately have that, for $g \in \ccal_b(\rbb_+)$,
\begin{equation}
\mean_a\big(f(\cdot)g(\lambda\,\cdot)\big) = \frac{1}{2}\int_{-1}^1 dx \,f(a^x)g(\lambda a^x) 
                                           = \frac{1}{2\log a}\int_{\rbb_+}\frac{d\mu}{\mu}\,f(\mu)g(\lambda \mu).
\end{equation}
(Note that the first integrand is a continuous function on $[-1,1]$.) This proves \ref{convolIntegral}.
\cmpqed\end{proof} 

We now use the above lemma to show that it is possible to smear elements of $\ubfk$ with respect to 
dilations.

\begin{lemma} \label{normContLemm}
Let $a>1$, and let $f\in \ccal_b(\rbb_+)$ with $\supp f \subset (1/a,a)$. There exists a bounded linear operator
$\udelta[f]:\ubfk \to \ubfk$ such that:
\begin{enumerate}
\renewcommand{\theenumi}{(\roman{enumi})}
\renewcommand{\labelenumi}{\theenumi}
\item $\|\udelta[f]\| \leq 2\|f\|_\infty \log a$;
\item \label{smearedLocal} if $\uB_\lambda \in \afk(\lambda\ocal)$ for all $\lambda > 0$, then 
$(\udelta[f]\uB)_\lambda \in \afk(\lambda \ocal_1)$ where $\ocal_1$ is any open bounded region such that
$\mu \ocal \subset \ocal_1$ for all $\mu \in [1/a,a]$, and if furthermore the function 
$(x,\Lambda) \in\poincare\mapsto\ualpha_{x,\Lambda}(\uB)$ is norm continuous, then 
$\udelta[f]\uB \in \uafk(\ocal_1)$;
\item if $\uB \in \uafk$, then $\udelta[f]\uB = \int_{\rbb_+}  f(\mu) \, \udelta_\mu(\uB) \, d\mu / \mu$ as
a Bochner integral.
\end{enumerate}
\end{lemma}

\begin{proof}
Let $\chi, \psi \in \hcal$ be arbitrary vectors. Given $\uB \in \ubfk$, consider the
function $g \in \bcal(\rbb_+)$ defined by $g(\lambda) = \hrskp{\chi}{\uB_\lambda\psi}$. Since $\|g\|_\infty
\leq \|\uB\|\|\chi\| \|\psi\|$, by (i) of the previous lemma the equation
\begin{equation}\label{defudelta}
\hrskp{\chi}{(\udelta[f]\uB)_\lambda\psi} = (\convol_f g)(\lambda)
\end{equation}
uniquely defines an element $\udelta[f]\uB \in \ubfk$, and property (i) is satisfied for
$\udelta[f]$. If furthermore $\uB \in \uafk$, then $g \in \ccal_b(\rbb_+)$, and therefore
(iii) follows from the analogous statement of the previous lemma. If now $\uB_\lambda \in \afk(\lambda \ocal)$
and $\ocal_1 \supset \mu \ocal$ for all $\mu \in [1/a,a]$, substituting $\hrskp{\chi}{\cdotarg\psi}$
with $\hrskp{\chi}{[A,\cdot]\psi}$, $A \in \afk(\lambda \ocal_1)'$,  in Eq.~\eqref{defudelta}
immediately entails $(\udelta[f]\uB)_\lambda \in \afk(\lambda \ocal_1)$. Suppose now that 
$(x,\Lambda) \in\poincare\mapsto\ualpha_{x,\Lambda}(\uB)$ is norm continuous. In order to show
that $\udelta[f]\uB \in \uafk(\ocal_1)$ it is now sufficient to show that the functions
$\mu \in \rbb_+ \mapsto \udelta_\mu(\udelta[f]\uB)$ and $(x,\Lambda)\in\poincare \mapsto
\ualpha_{x,\Lambda}(\udelta[f]\uB)$ are norm continuous at the identity of
the respective groups. Since $\hrskp{\chi}{\udelta_\mu(\udelta[f]\uB)_\lambda\psi}=(\convol_f g)(\mu\lambda)$,
continuity with respect to dilations follows at once from the estimate~\eqref{eq:dilationcont}.
For Poincar\'e transformations, we proceed as follows. For $\kappa \in \rbb_+$, set
\begin{equation}
  g_{x,\Lambda,\kappa}(\lambda) := \bighrskp{U(\kappa x,\Lambda)\st\chi}{ \uB_\lambda U(\kappa x,\Lambda)\st \psi}.
\end{equation}
With this definition, we have
\begin{equation}
  \bighrskp{\chi}{\big( (\ualpha_{x,\Lambda}\udelta[f]\uB)_\lambda - (\udelta[f]\uB)_\lambda\big)\psi}
  = \big( \convol_f(g_{x,\Lambda,\lambda}-g) \big) (\lambda).
\end{equation}
Estimating by Lemma~\ref{lem:meanconvolution} \ref{convolNorm}, we obtain after a straightforward computation,
\begin{equation}
  \| \ualpha_{x,\Lambda}\udelta[f]\uB  - \udelta[f]\uB \|
  \leq 2 \log a\: \|f\|_\infty \sup_{\mu\in [1/a,a]} \| \ualpha[\mu^{-1}x,\Lambda] \uB-\uB \|.
\end{equation}
This vanishes as $(x,\Lambda)\to\id$, since Poincar\'e transformations act norm-continuous
on $\uB$ by assumption; thus \ref{smearedLocal} is proved.
\cmpqed\end{proof}

Since elements $\uB \in \ubfk$ such that $(x,\Lambda)\mapsto\ualpha_{x,\Lambda}(\uB)$ is norm
continuous can be easily constructed by smearing in the traditional way over the
Poincar\'e group \cite{BucVer:scaling_algebras}, 
the above lemma shows the existence of a large family of elements satisfying
the continuity conditions imposed in the definition of the scaling algebra $\uafk$.

\subsection{Scaling limit}

The scaling limit of the theory is defined by the limits of our operator-valued scaling functions in the vacuum state.
Consider the following states $\uomega_\lambda$ on $\uafk$:
\begin{equation}
  \uomega_\lambda : \; \uA \mapsto \omega(\uA_\lambda).
\end{equation}
The rough idea for constructing a scaling limit theory is to take the limit of these states as $\lambda \to 0$, and then to consider the GNS representation of $\uafk$ with respect to this limit state.
However, while the above expression may converge for certain operators $\uA$ in relevant
examples \cite{BucVer:scaling_examples}, the limit will certainly not exist in general.

The approach taken in \cite{BucVer:scaling_algebras} is to choose a
weak-$\ast$ cluster point of the set $\{ \uomega_\lambda \}$, which exists by
the Alaoglu-Bourbaki theorem. These states were shown to be Poincaré invariant pure vacuum states.

We will use a somewhat more general approach here:
Let $\mean$ be a mean on the semigroup $(0,1]$, with multiplication. That is,
$\mean$ is a linear functional on $\bcal((0,1])$, such that
$\mean(\idop)=1$, and $\mean(f) \geq 0$ for $f \geq 0$. 
The functional is automatically bounded by $|\mean(f)|\leq \sup_\lambda|f(\lambda)|$.
We now define\footnote{%
For simplicity of notation, we will often write $\mean(f(\lambda))$
as a shorthand for $\mean(\lambda \mapsto f(\lambda))$.
}
a functional $\uomega$ on the scaling algebra as
\begin{equation} \label{uomegaDef}
  \uomega (\uA) := \mean(\omega(\uA_\lambda)), \quad \uA \in \uafk.
\end{equation}
Then $\uomega$ is a linear, positive, normalized functional on $\uafk$, hence a state. 
This construction covers in particular the following relevant cases:
\begin{enumerate}
\renewcommand{\theenumi}{(\alph{enumi})}
\renewcommand{\labelenumi}{\theenumi}
\item \label{mFixedscale}
The mean $\mean$ is an \emph{evaluation functional}, i.e.~$\mean(f)=f(\lambda_0)$ for some fixed $\lambda_0$.
Then $\uomega$ is the ``vacuum at scale $\lambda_0$'', and the GNS representation corresponds to
the theory at this scale.
\item \label{mLimit}
$\mean$ is a \wstar limit point of such evaluation functionals as $\lambda_0 \to 0$. These are
the scaling limit states considered by Buchholz and Verch in \cite{BucVer:scaling_algebras}.
\item \label{mInvariant}
$\mean$ is an \emph{invariant mean} on the semigroup; that is, $\mean(f_\mu) = \mean(f)$, 
where $f_\mu(\lambda) = f(\mu\lambda)$,
$0 < \mu \leq 1$. It is well known that such invariant means exist, although they are not unique.
This gives an alternative version of the scaling limit theory, which we will investigate
in more detail below.
\end{enumerate}

Cases \ref{mFixedscale} and \ref{mLimit} share the property that $\mean$ is multiplicative,
i.e.~$\mean(fg) = \mean(f) \mean(g)$. In fact, it is known \cite[Ch.~IV.6]{DunSch:linop1} that \wstar limit points of evaluation functionals are the only multiplicative means on $(0,1]$.
On the other hand, $\mean$ is precisely \emph{not} multiplicative in case \ref{mInvariant},
since there can be no multiplicative invariant means on nontrivial abelian semigroups \cite{Mit:mult_invar_means}.

Further, cases \ref{mLimit} and \ref{mInvariant} are \emph{asymptotic} means, in the sense that $\mean(f) = 0$
whenever $f$ vanishes on a neighborhood of 0. Such asymptotic means are generalizations of the limit $\lambda \to 0$: 
Namely, if $f(\lambda)$ converges as $\lambda \to 0$, it follows that $\mean(f)=\lim_{\lambda}f(\lambda)$.
Also, along the line of ideas given in \cite[Corollary~4.2]{BucVer:scaling_algebras}, one can show that if
$\mean$ is asymptotic, the vacuum state $\omega$ in Eq.~\eqref{uomegaDef} can be replaced with any other
locally normal state in the original theory, without changing the resulting state $\uomega$ on the scaling algebra. 

For the following, we will usually choose a fixed mean $\mean$. Since we are mainly
interested in asymptotic means,\footnote{%
However, most of our results do not rely on this property; they apply to other means $\mean$ as well, and are not limited
to  cases \ref{mLimit} and \ref{mInvariant}.}
we will refer to the corresponding state on $\uafk$
as the \emph{scaling limit state} $\uomega_0$. The scaling limit theory
is now obtained by a GNS construction with respect to this state. 
We denote this GNS representation of $\uafk$ as $\pi_0$, and the representation Hilbert
space as $\hcal_0$, where $\Omega_0 \in \hcal_0$ is the GNS vector. 

We can also transfer the symmetry group action to the representation space $\hcal_0$; but here the properties of $\mean$
are crucial. We first note:
\begin{lemma}
One has $\uomega_0 \circ \ualpha_g = \uomega_0$ for all Poincaré transformations $g=(1,x,\Lambda)$.
If $\mean$ is invariant, the same holds for all $g \in \ugcal$.
\end{lemma}
\begin{proof}
For Poincaré transformations, $\uomega_0 \circ \ualpha_g = \uomega_0$ follows from the invariance
of $\omega$ under $\alpha_{x,\Lambda}$ at finite scales, and for dilations it follows from the invariance of $\mean$.
\cmpqed\end{proof}

Now, in the limit theory, we can implement the subgroup of those symmetries that leave $\uomega_0$ invariant:

\begin{theorem} \label{limitSpectrumThm}
Let $\uomega_0$ be a scaling limit state, and let $\gcal_0\subset \ugcal$ be the subgroup of all $g$ which fulfill 
$\uomega_0 \circ \ualpha_g = \uomega_0$. 
There exists a strongly continuous unitary representation $U_0$ of $\gcal_0$ on $\hcal_0$,
such that $\alpha_{0,g} \circ \pi_0 = \pi_0 \circ \ualpha_g$ with $\alpha_{0,g} = \ad U_0(g)$, for all $g \in \gcal_0$.
One has $U_0(g)\Omega_0 = \Omega_0$ for $g \in \gcal_0$. The representation $U_0(x)$ of translations fulfills the spectrum condition.
\end{theorem}

\begin{proof}
We set
\begin{equation}
  U_0(g) \pi_0(\uA) \Omega_0 := \pi_0(\ualpha_g \uA) \Omega_0.
\end{equation}
This defines $U_0(g)$ on a dense set; it is well-defined, since $\ker \pi_0$
is invariant under $\ualpha_g$ by assumption. One easily checks that $U_0(g)$ 
is norm-preserving, and thus can be extended to a unitary on all of $\hcal_0$.
Strong continuity of the representation follows from norm continuity of
$g \mapsto \ualpha_g \uA$. The properties $U_0(g)\Omega_0 = \Omega_0$ 
and $\alpha_{0,g} \circ \pi_0 = \pi_0 \circ \ualpha_g$ are immediate.
For the spectrum condition, it suffices to show that
\begin{equation}
  \int dx \, f(x) \, \bighrskp{\pi_0(\uA) \Omega_0}{U_0(x) \pi_0(\uA') \Omega_0} = 0 
\end{equation}
for all $\uA,\uA' \in \uafk$, and all test functions $f \in \scal(\rbb^{s+1})$ such that
the Fourier transform of $f$ vanishes on the closed forward lightcone.
Due to norm continuity of the representation, we can rewrite this expression as
\begin{equation}
  \int dx \, f(x) \, \bighrskp{\pi_0(\uA) \Omega_0}{U_0(x) \pi_0(\uA') \Omega_0} 
% = \uomega_0(\uA\st \ualpha_f(\uA') \big)
 = \mean \Big( \int dx \, f(x) \, \omega(\uA^\ast_\lambda \alpha_{\lambda x} \uA'_\lambda) \, \Big).
\end{equation}
But the right-hand side vanishes due to the spectrum condition in the original theory $\afk$.
This concludes the proof.
\cmpqed\end{proof}

We now define the local net in the limit theory as 
$\afk_0(\ocal) := \pi_0(\uafk(\ocal))''$.
It is clear that $\afk_0$ is local, isotone, and covariant under $\alpha_{0,g}$, since these
properties transfer from $\uafk$ on an ultraweakly dense set. 
By the above results, we have established $\afk_0$ as a local net of algebras in a positive energy representation with symmetry group $\gcal_0$,
in the sense of Definition~\ref{netDef}.

Now as a last and crucial point, we ask whether the limit vacuum is a pure state, 
or (equivalently) $\Omega_0$ is unique as a translation-invariant vector, or (equivalently) the representation $\pi_0$ is irreducible.
For the case of \emph{multiplicative} means $\mean$ in $s \geq 2$ dimensions,\footnote{%
In $1+1$ space-time dimensions, the analogue is false: Even if $\mean$ is 
multiplicative, $\pi_0(\uafk)$ may contain a nontrivial center. 
An example for this behavior occurs in the Schwinger model \cite{BucVer:scaling_examples}.
However, since the phase space conditions we use in Sec.~\ref{fieldSec} do not apply
to this class of models a priori, we do not place emphasis on this situation here.}
a positive answer has been given in \cite{BucVer:scaling_algebras}. 

For non-multiplicative $\mean$, and in particular if $\mean$ is invariant,
the same is however impossible: Here already $\pi_0 \restrict \zfk(\uafk)$ is known to be reducible,
and the space of translation-invariant vectors must be more than 1-dimensional.
The structure of the limit theory may be more complicated in this case.
Since it is not directly relevant to our current line of arguments, we will confine ourselves
to some remarks here, leaving the details -- which are of interest in their own right -- 
to a separate discussion \cite{BDM:scaling2}.

First, it can be shown that the nontrivial image of the center $\zfk(\uafk)$ is the
only ``source'' of reducibility: namely one has $\pi_0(\uafk)' = \pi_0(\zfk(\uafk))''$ if $s \geq 2$.
Here the case of a vacuum limit state arises as a special case for $\pi_0(\zfk(\uafk)) = \cbb\idop$.
For more general limit states, one would like to decompose the limit net $\afk_0$ along its center.
By the representation theory of the \emph{commutative} $C\st$ algebra $\zfk(\uafk)$, one has a 
canonical decomposition
\begin{equation} \label{algebraDecomp}
  \uomega_0(\uA) = \int_Z d\nu(z)\; \uomega_z(\uA), \quad
  \uA \in \uafk,
\end{equation}
where $Z$ is a compact Hausdorff space, $\nu$ a regular Borel measure on $Z$, and $\uomega_z$ are
scaling limit states that correspond to \emph{multiplicative} means. One would naturally want to interpret
Eq.~\eqref{algebraDecomp} in terms of a direct integral of Hilbert spaces, and decompose the representation $\pi_0$
into corresponding irreducible representations $\pi_z$. This faces technical problems however:
In particular for invariant means $\mean$, the limit Hilbert space $\hcal_0$ is not
separable \cite{dou:invar_mean}, and so the standard methods of decomposition theory
cannot be applied (see e.g.~\cite{DriSum:central_decomp}, \cite[Ch.~14]{KadRin:algebras2}).
One needs to use generalized notions of direct integrals for nonseparable spaces
\cite{Wil:direct_integrals_1,Sch:direct_integral_repr}. We do not enter this discussion here.
Let us just note that for the case of a unique vacuum structure, as introduced in \cite{BucVer:scaling_algebras},
and under additional regularity assumptions, it can be shown that the limit theory $\afk_0$
has a simple product structure:
\begin{equation} \label{limitTens}
    \afk_0(\ocal) \isom \pi_0(\zfk(\uafk))'' \wtens \hat\afk(\ocal),
\end{equation}
where $\hat\afk$ is a fixed local net in a vacuum sector, independent of
$\mean$. For a free real scalar field of mass $m$ in physical space-time, the
net $\hat\afk$ would correspond to a massless free field. Note that the factor $\pi_0(\zfk(\uafk))''$ depends on the mean $\mean$, but not on the specific theory $\afk$ 
in question. The symmetry group operators $U_0(g)$ also factorize along the above product:
One has $U_0(g) \isom (U_0(g)\restrict \hcal_\zfk) \otimes \hat U(g)$, where $\hcal_\zfk$ is 
the Hilbert space generated by $\pi_0(\zfk(\uafk))$, and $\hat U$ the representation associated
with $\hat\afk$. 
In the case of an invariant mean, this representation includes the dilations. 
But while for Poincaré transformations $U_0(g)\restrict \hcal_\zfk$ is trivial,
dilations have a nontrivial action on $\hcal_\zfk$.

To summarize: If $\mean$ is multiplicative, in particular in case \ref{mLimit} above, we obtain a theory
in the vacuum sector, with a pure vacuum state. It need not be dilation covariant, however.
If $\mean$ is invariant, i.e.~in case \ref{mInvariant}, 
dilations can canonically be implemented in the limit; but the limit state $\uomega_0$ is not pure. 

\subsection{States and energy bounds in the limit} \label{energyBoundsSec}

So far, we have described the theory on the level of bounded observables, at which the scaling limit
can be computed. For the analysis of pointlike quantum fields, however, we need to consider
a more general structure, which is tied to the Hilbert space representations of the scaling algebra.
In this section, we will not yet refer to the locality properties of pointlike fields, but only
be concerned with their singular high energy behavior.

We first describe the situation in the original theory $\afk$. Here, let $\Sigma = \boundedops_\ast$
be the predual Banach space of $\boundedops$, i.e.~the set of normal functionals. We are interested
in functionals with an energy cutoff $E$, and therefore define
\begin{equation}
 \Sigma(E) := \bigsetprop{ \sigma (P(E) \cdotarg P(E)) }{\sigma \in \Sigma}.
\end{equation}
Setting $R := (1+H)^{-1}$, a bounded operator, we can also consider the space of smooth normal functionals:
\begin{equation}
 \cinftys := \bigsetprop{ \sigma \in \Sigma }{ \| \sigma(R^{-\ell}\cdotarg R^{-\ell}) \|<\infty \text{ for all }\ell>0}.
\end{equation}
We equip this space with the Fréchet topology induced by all the norms 
$\lnorm{\sigma}{\ell} := \|\sigma ( R^{-\ell} \cdotarg R^{-\ell} )  \|$, $\ell > 0$.
Its dual space in this topology is 
\begin{equation}
 \cinftyss = \bigsetprop{ \phi : \cinftys \to \cbb }{ \lnorm{\phi}{\ell} := \| R^{\ell}\phi R^\ell \|<\infty \text{ for some }\ell>0}.
\end{equation}
This is the space in which we expect our pointlike fields $\phi(x)$ to be contained.
The ``polynomial energy damping'' with powers of $R$ plays an important role in our analysis, and we will often need the
following key lemma; cf. \cite[Lemma 3.27]{Bos:operatorprodukte}.
\begin{lemma}\label{polyBoundsLemm}
For any $\ell > 0$ there exists a constant $c_\ell>0$ with the following property:
Let $\hcal$ be a Hilbert space, and $H \geq 0$ a positive selfadjoint operator, possibly unbounded, on a dense
domain in $\hcal$. Let $P(E)$ be its spectral projections, and $R = (1+H)^{-1}$.
If $c>0$, and $\phi$ is a sesquilinear form on a dense set of $\hcal\times\hcal$
such that
\begin{displaymath}
 \| P(E) \,\phi \,P(E) \| \leq c \cdot (1+E)^{\ell-1} \quad \forall E > 0,
\end{displaymath}
then it follows that
\begin{displaymath}
\| R^{\ell}  \,\phi\,  R^{\ell} \| \leq c_\ell  c .
\end{displaymath}
\end{lemma}
It is important here that $c_\ell$ depends on $\ell$ only, not on $H$, $\phi$, or $c$.

\begin{proof}
The spectral theorem applied to $R^\ell$ yields for any $\chi \in \hcal$,
\begin{equation}
 \hrskp{\chi}{R^\ell\chi} = \int (1+E)^{-\ell} d\hrskp{\chi}{P(E) \chi}.
\end{equation}
Integrating by parts in this Lebesgue-Stieltjes integral \cite[Ch.~III Thm.~(14.1)]{Sak:integration}, we obtain the following formula in the sense of matrix elements:
\begin{equation} \label{rFormula}
  R^\ell = \ell \int_0^\infty dE \, (1+E)^{-\ell-1} \,P(E).
\end{equation}
Now let $E>0$ be fixed, and let $\chi,\chi' \in P(E)\hcal$ be unit vectors. 
Using Eq.~\eqref{rFormula} twice, we obtain
\begin{multline}
| \hrskp{\chi\, }{ R^\ell\,  \phi\,  R^\ell\; \chi' } |
=   \ell^2 \Big| \int  dE_1 dE_2\;
  \frac{   \hrskp{\chi \, }{ \, P(E_1) \, \phi\,  P(E_2)\,  | \, \chi' } } 
{ (1\!+\!E_1)^{\ell+1} (1\!+\!E_2)^{\ell+1}  } \Big|
\\
\leq c \cdot
\ell^2 \int \limits_0^\infty dE_1 \int \limits_0^\infty dE_2
 \frac{(1 + \max\{E_1,E_2\})^{\ell-1}}{(1\!+\!E_1)^{\ell+1} (1\!+\!E_2)^{\ell+1} }\;.
\end{multline}
In the estimate, we have used the hypothesis of the lemma. 
The integral on the right-hand side exists, since the integrand vanishes like $E_{i}^{-2}$
in both variables; and this bound is independent of $E$.
This allows us to extend $R^\ell \phi R^\ell$ to a bounded operator, and
gives us an estimate for its norm that depends only on $\ell$.
\cmpqed\end{proof}

In the limit theory, we use analogous definitions for the spaces $\Sigma_0$, $\Sigma_0(E)$, 
$\cinfty(\Sigma_0)$, and $\cinfty(\Sigma_0)\st$, referring to the Hamiltonian $H_0$
and its spectral projectors $P_0(E)$. Note that Lemma~\ref{polyBoundsLemm} above
holds true for $H_0$ in place of $H$ as well.

It is not obvious however how to obtain corresponding structures on the scaling algebra $\uafk$,
in order to describe the limiting procedure. Difficulties arise because the representation $\ualpha$
is not unitarily implemented, and hence we cannot refer to its generators and their spectral projections:
an energy operator on the scaling algebra is not available.
However, it is possible to consider operators in $\uafk$ of finite energy-momentum \emph{transfer}.
We describe them here as follows:\footnote{One can more abstractly define the spectral support of
an operator $\uA \in \uafk$ with respect to translations, and define the space $\tilde\uafk(E)$
by this means. We do not need the general formalism here, however.}
\begin{multline}
  \tilde\uafk(E) := \bigsetprop{ \ualpha_f \uA = \int d^{s+1}x \, f(x) \ualpha_x \uA \; }{ 
\\ \;
	\uA \in \uafk, \, f \in \scal(\mkr), \, \supp \tilde f \subset (-E,E)^{s+1}}.
\end{multline}
Here we refer to the translation subgroup of $\ualpha$ only. It is clear that
$\tilde\uafk(E)$ is a linear space, closed under the $\ast$ operation. Moreover,
for any fixed $g \in \ugcal$, we have $\ualpha_g \tilde\uafk(E) \subset \tilde\uafk(E')$
for suitable $E'$.

The important point is now that the representors of $\uA \in \tilde\uafk(E)$ generate
energy-bounded vectors from the vacuum, with the correct renormalization. To formulate
this, we consider in addition to $P(E)$ also $P(E-0)$, the spectral projector
of $H$ for the interval $(-\infty,E)$; it differs from $P(E)$ by the 
projector onto the eigenspace\footnote{%
It is in fact not expected that $H$ has any eigenvectors other than $\Omega$,
at least under reasonable assumptions on the phase space behavior of the theory \cite{Dyb:unique_vacuum}. 
We do however not rely on this property.}
with eigenvalue $E$.
Correspondingly, we use $P_0(E-0)$ in the limit theory.

\begin{proposition} \label{energyDenseProp}
For any $\uB \in \tilde\uafk(E)$, one has
$\uB_\lambda \Omega \in P(E/\lambda-0) \hcal$ and $\pi_0(\uB) \Omega_0 \in P_0(E-0) \hcal_0$.
Further, the inclusion
%\begin{equation*}
   $\pi_0(\tilde \uafk(E)) \Omega_0 \subset P_0(E-0) \hcal_0 $
%\end{equation*} 
is dense.
\end{proposition}
\begin{proof}
  Let $\chi \in \hcal$, and let $\uB = \ualpha_f \uA \in \tilde\uafk(E)$. With $Q_\kappa$ being
the spectral projectors of the momentum operators $P^\kappa$, we compute
\begin{equation}
  \hrskp{\chi}{\uB_\lambda \Omega} = \int d^{s+1}x \,f(x) \hrskp{\chi}{U(\lambda x) \uA_\lambda \Omega}
= \int \tilde f(p) d^{s+1}\hrskp{\chi}{\prod_\kappa Q_\kappa(p^\kappa/\lambda) \uA_\lambda \Omega}.
\end{equation}
Now if $\chi \in (1-P(E/\lambda))\hcal$, or if $\chi$ is an eigenvector of $H$ with eigenvalue
$E/\lambda$, then the right-hand side vanishes by the support properties of $\tilde f$.
This shows $\uB_\lambda \Omega \in P(E/\lambda-0) \hcal$. The proof for $\pi_0(\uB) \Omega_0$
is analogous.

Now suppose that the inclusion $\pi_0(\tilde \uafk(E)) \Omega_0 \subset P_0(E-0) \hcal_0 $ was not dense. 
Then we can find $\chi \in P_0(E-0) \hcal_0$,
$\chi \neq 0$, such that
\begin{equation} \label{nondenseAssumpt}
   \hrskp{\chi}{ \pi_0(\uB) \Omega_0 } = 0
   \quad \text{for all }
   \uB \in \tilde \uafk(E).
\end{equation} 
This means that, whenever $f$ is a test function with $\supp \tilde f \subset (-E,E)^{s+1}$, we have
\begin{equation} 
  0 = \int d^{s+1} x \, f(x)  \hrskp{\chi}{ U_0(x) \pi_0(\uA) \Omega_0 } 
    = \int_{\bar \vcal_+} \tilde f(p) d^{s+1} \nu(p)
   \quad \text{for all }
   \uA \in \uafk,
\end{equation} 
where $\nu(p)$ is a measure, the Fourier transform of $\hrskp{\chi}{ U_0(x) \pi_0(\uA) \Omega_0 }$.
We specifically choose $f(x) = f_T(x^0) f_S(\xv)$, where $\supp \tilde f_T \subset (-E,E)$, 
$\supp \tilde f_S \subset (-E,E)^s$. By the spectrum condition, the support of $\nu(p)$ is within the 
closed forward light cone; thus we can actually remove the constraints on $\supp \tilde f_S$.
Choosing for $f_S$ a delta sequence in configuration space, we obtain that 
\begin{equation} 
  0 = \int \tilde f_T(p_0)\,d\nu_T(p_0)
   \quad \text{whenever } \supp \tilde f_T \subset (-E,E).
\end{equation} 
Here $\nu_T(p_0)$ is the Fourier transform of $\hrskp{\chi}{ U_0(t) \pi_0(\uA) \Omega_0 }$, referring to 
time translations only. Thus the measure $\nu_T$ must have its support in $(-E,E)^c$.
On the other hand, the spectrum condition implies that $\supp \nu_T \subset [0,E]$.
Hence $\supp \nu_T = \{E\}$, and $\nu_T$ is a delta measure: $\nu_T = c \, \delta(p_0-E)$
with some constant $c$. 
By Fourier transformation, that yields
\begin{equation} 
   \hrskp{\chi}{U(t) \pi_0(\uA) \Omega_0} = \hrskp{\chi}{\pi_0(\uA) \Omega_0} e^{iEt}
   \text{ for all $\uA \in \uafk$, $t \in \rbb$.}
\end{equation}
Since $\pi_0(\uafk)\Omega_0$ is dense in $\hcal_0$, this means that
$\chi$ is an eigenvector of $H_0$ with eigenvalue $E$. 
But this contradicts $\chi \in P_0(E-0)\hcal_0$.
Thus the said inclusion must be dense.
\cmpqed\end{proof}

We will now investigate in more detail the convergence of states in the limit.
The heuristic picture is that functionals like
\begin{equation}
   \sigma_\lambda = \mtxe{\uB_\lambda \Omega}{\cdotarg}{\uB_\lambda' \Omega}, \quad
   \uB,\uB' \in \tilde\uafk(E),
\end{equation}
``converge'' to an energy-bounded limit state as $\lambda \to 0$. This can easily
be understood if evaluating them on bounded operators $\uA \in \uafk$; we will however
need the same also for unbounded objects. Let us formalize this more strictly. We set
\begin{equation}
  \uSigma(E) := \bigsetprop{ \usigma : \rbb_+ \to \Sigma }{ \usigma_\lambda(A) = \omega(\uB_\lambda\st A \uB_\lambda') 
  \text{ for some } \uB,\uB' \in \tilde\uafk(E)}.
\end{equation}
We also define $\uSigma = \cup_{E>0} \uSigma(E)$. These sets are not linear spaces,
but this will not be needed for our purposes. Note that, via the action on $\uB$ and $\uB'$,
we have a natural action $\ualpha\st_g$ of $\ugcal$ on $\uSigma$, which fulfills
\begin{equation}
   (\ualpha_g\st \usigma)_\lambda ((\ualpha_g\uA)_\lambda) =
   \usigma_{\mu\lambda}(\uA_{\mu\lambda}), \quad \text{where}\;
   \usigma \in \uSigma, \, \uA \in \uafk, \, g = (\mu,x,\Lambda) \in \ugcal.
\end{equation}
We now define the scaling limit of $\usigma \in \uSigma$, denoted by $\pi_0\st \usigma \in \Sigma_0$, via
\begin{equation}
  \pi_0\st \usigma (\pi_0 \uA) := \mean( \usigma_\lambda (\uA_\lambda) ) = \uomega_0 (\uB\st \uA \;\uB' ).
\end{equation}
It is clear that this is well-defined. By Proposition~\ref{energyDenseProp} above, the span of all
$\pi_0\st \usigma$, $\usigma \in \uSigma(E)$, is dense in $\Sigma_0(E-0)$, and the union over all
$E>0$ is dense in $\cinfty(\Sigma_0)$ in the corresponding topology. 
We also have $\pi_0\st ( \ualpha_g\st \usigma) = (\pi_0\st \usigma) \circ  \alpha_{0,g}^{-1}$
in a natural way. Let us further note:

\begin{lemma} \label{sigmaNormLemm}
For all $\usigma \in \uSigma$, it holds that $\mean (\|\usigma_\lambda\|) \leq \| \pi_0\st \usigma \|$.
\end{lemma}

\begin{proof} 
First, it is clear that we have $\|\pi_0(\uA)\Omega_0\|^2 = \mean(\|\uA_\lambda\Omega\|^2)$ for any $\uA\in\uafk$.
Now let $\usigma \in \uSigma$ with $\usigma_\lambda = \mtxe{\uB_\lambda \Omega}{\cdotarg}{ \uB_\lambda' \Omega}$.
The mean $\mean$, as a state on a commutative algebra, satisfies the Cauchy-Schwarz inequality,
from which we can conclude:
\begin{multline}
 \mean(\|\usigma_\lambda\|) = \mean (\|\uB_\lambda \Omega\|\|\uB_\lambda' \Omega\|)
 \leq \mean (\|\uB_\lambda \Omega\|^2)^{1/2}  \;\mean (\|\uB_\lambda' \Omega\|^2)^{1/2}
\\
 = \|\pi_0(\uB) \Omega_0\| \|\pi_0(\uB') \Omega_0\| = \|\pi_0\st \usigma\|.
\end{multline}
This proves the lemma.
\cmpqed\end{proof}

We can now use this structure to describe the scaling limit behavior of unbounded objects,
more general than the bounded operator sequences $\uA \in \uafk$. Namely, we consider
functions $\lambda \mapsto \uphi_\lambda$, with values in $\cinftyss$. These will later 
be sequences of pointlike fields with renormalization factors. Here we are only interested
in their high-energy behavior. We use a notion of ``uniform'' polynomial energy damping
at all scales: For $\lambda>0$, set $\uR_\lambda := (1+\lambda H)^{-1}$. 
This $\uR$ is an element of $\ubfk$, but not of $\uafk$. 
We can, however, multiply $\uphi$ with powers of $\uR$ from the left or right,
this product being understood ``pointwise''. We can then consider the norms
\begin{equation}
          \lnorm{\uphi}{\ell} = \sup_\lambda \|\uR_\lambda^\ell \uphi_\lambda \uR_\lambda^\ell\|
\end{equation} 
on the spaces of those functions $\uphi$ where the supremum is finite. 
We also have a notion of symmetry transformation on the functions $\uphi$,
defined in a natural way as 
\begin{equation}
  (\ualpha_{\mu,x,\Lambda} \uphi)_\lambda := U(\mu\lambda x,\Lambda) \uphi_{\lambda \mu} U(\mu\lambda x,\Lambda)\st.
\end{equation}
Our space of ``regular'' $\uphi$ is now defined as follows, in analogy to the algebras $\uafk$.
\begin{equation} \label{phiSpaceDef}
\uPhi := \bigsetprop{ \uphi : \rbb_+ \to \cinftyss }{ \exists \ell: \; \lnorm{\uphi}{\ell} < \infty; \;
	 g \mapsto \ualpha_g \uphi \text{ is continuous in } \lnorm{\cdotarg}{\ell} }.
\end{equation}
By the natural inclusion $\boundedops \hookrightarrow \cinftyss$, we have an embedding
 $\uafk \hookrightarrow \uPhi$, compatible with the symmetry action.
We note that the continuity requirement in Eq.~\eqref{phiSpaceDef} is trivially fulfilled for the translation subgroup,
since $\| \lambda P^\mu \uR_\lambda   \| \leq 1$ at all scales by the spectrum condition. 
For Lorentz transformations
and dilations, the requirement is nontrivial; it suffices however to check continuity at $g=\id$,
as is easily verified using the commutation relations between $\ualpha_g$ and $\uR$.

The functions in $\uPhi$ are sufficiently regular to allow the definition of a scaling limit. Namely, we have:

\begin{proposition} \label{phiLimitProp}
Let $\uphi \in \uPhi$. There exists a unique element $\pi_0\uphi \in \cinfty(\Sigma_0)\st$
such that
\begin{equation*}
  (\pi_0\st \usigma)( \pi_0 \uphi) = \mean(\usigma_\lambda(\uphi_\lambda)) \quad
 \text{for all } \usigma \in \uSigma.
\end{equation*}
The map $\uphi \mapsto \pi_0\uphi$ is linear, and one has for anz $E>0$,
\begin{equation*}
  \| \pi_0 \uphi \restrict \Sigma_0 (E-0) \| \leq \sup_{0<\lambda\leq 1} \| \uphi_\lambda \restrict \Sigma(E/\lambda) \|.
\end{equation*}
For fixed $\ell$, there is a constant $c$ such that
%\begin{equation*}
  $\lnorm{ \pi_0 \uphi }{2\ell+1} \leq c \lnorm{ \uphi }{ \ell }$
%\end{equation*}
for all $\uphi$ for which the right-hand side is finite.
\end{proposition}

\begin{proof}
Again, let $\usigma_\lambda(\cdotarg) = \mtxe{\uB_\lambda \Omega}{\cdotarg}{ \uB_\lambda' \Omega}$,
with $\uB,\uB' \in \tilde\uafk(E)$, where $E>0$ is fixed in the following. 
Set  $\|\uphi\|_E := \sup_{0<\lambda\leq 1} \| \uphi_\lambda \restrict \Sigma(E/\lambda) \|$. 
We have
\begin{equation}
  | \usigma_\lambda(\uphi_\lambda) | \leq  \|\usigma_\lambda\| \|\uphi\|_E
\leq \|\uB\| \, \|\uB'\| \,  \|\uphi\|_E,
\end{equation}
which is a uniform estimate in $\lambda$. So we can apply the mean $\mean$ and 
obtain by Lemma~\ref{sigmaNormLemm}:
\begin{equation}
  | \mean(\usigma_\lambda(\uphi_\lambda)) | \leq \|\pi_0\st \usigma\| \|\uphi\|_E.
\end{equation}
This allows us to define $\pi_0\uphi$ on all functionals of the form
$\mtxe{\pi_0(\uB)\Omega_0}{\cdotarg}{\pi_0(\uB')\Omega_0}$. Using the
density properties outlined in Proposition~\ref{energyDenseProp}, $\pi_0\uphi$  can uniquely 
be extended to a bounded sesquilinear form on $P_0(E-0)\hcal_0 \times P_0(E-0)\hcal_0$,
and then to a linear form on $\Sigma_0(E-0):= P(E-0)\Sigma P(E-0)$ with the bounds
\begin{equation}
   \| \pi_0\uphi \restrict \Sigma_0(E-0) \| \leq \|\uphi\|_E.
\end{equation}
[Of course, the definition of the linear form at fixed $E$ is compatible with the inclusions
$\Sigma_0(E-0) \subset \Sigma_0(E'-0)$.] The estimate on the $\ell$-norms now
follows by applying 
Lemma~\ref{polyBoundsLemm} with respect to the operator $\lambda H$.
\cmpqed\end{proof}

We note that the map $\pi_0$ on $\uPhi$ defined above
is an extension of the representation $\pi_0$ of the algebra $\uafk$, i.e.,
it is compatible with the inclusion $\uafk \hookrightarrow \uPhi$. 
Using the action of $\ugcal$ on $\uSigma$, we also obtain that $\pi_0$ is
compatible with the action of the unitaries on $\hcal_0$:
\begin{equation} \label{phiLimitSymm}
   \alpha_{0,g} (\pi_0\uphi) = \pi_0 (\ualpha_g \uphi) \quad
   \text{for all } g \in \gcal_0.
\end{equation}

Let us once more return to the operators $\uR$,
which play an important role since they have known commutation relations with 
``smeared'' sequences from $\uPhi$. Let $f \in \scal(\mkr)$ be a test function on Minkowski space, and set
\begin{equation} \label{smearedField}
  \ualpha_f \uphi := \int d^{s+1}x\,f(x) \ualpha_x \uphi
\end{equation}
as a weak integral, referring to the translation subgroup only.
It is known from \cite{FreHer:pointlike_fields} that $(\ualpha_f \uphi)_\lambda$ are more regular than
linear forms on $\cinftys$: They are actually unbounded operators on $\cinftyh = \cap_{\ell>0}R^\ell \hcal$.
In generalization of \cite[Eq.~2.4]{FreHer:pointlike_fields}, one easily proves the following
relation, valid at fixed $\lambda$ in the sense of matrix elements:
\begin{equation} \label{phiRComm}
  [ \uR , \ualpha_f \uphi] = - i \uR (\ualpha[\partial_0 f] \uphi) \uR.
\end{equation}
This will be crucial in our approximation of pointlike fields later.

As a last point, let us consider a finite-dimensional subspace $\ecal \subset \cinftyss$. 
It is well-known that $\ecal$ is always complementable in the locally convex space $\cinftyss$;
i.e.~there exists a continuous projection $p$ onto $\ecal$. We will need the fact that
such projections can be chosen uniformly at all scales. To that end, we set
$\uPhi^{(\ell)} := \{ \uphi \in \uPhi \, | \, \lnorm{ \uphi }{\ell}< \infty \}$.

\begin{proposition}  \label{uniformProjProp}
Let $\ecal \subset \cinftyss$ be a finite-dimensional subspace such that $\alpha_\Lambda \ecal = \ecal$
for all Lorentz transforms $\Lambda$. Let $\ell>0$ be large enough such that
$\lnorm{\phi}{\ell}<\infty$ for all $\phi \in \ecal$. There exists a map $\up: \uPhi^{(\ell)} \to \uPhi^{(\ell)}$
of the form $(\up \,\uphi)_\lambda = p_\lambda \uphi_\lambda$, where each $p_\lambda$ is projector onto $\ecal$,
and a constant $c>0$, such that $\lnorm{\up \,\uphi}{\ell} \leq c \lnorm{\uphi}{\ell}$.
\end{proposition}
We will refer to the map $\up$ (for given $\ecal$ and $\ell$) as a \emph{uniform projector}
onto $\ecal$.

\begin{proof}
Let $\ecal_\lambda$ be the space $\ecal$ equipped with the norm 
$\|\cdotarg\|_\lambda = \|\uR_\lambda^\ell \cdotarg \uR_\lambda^\ell \|$,
and let $\fcal_\lambda$ be $\{ \phi \in \cinftyss \,|\, \lnorm{\phi}{\ell} < \infty \}$ with the same norm. Then there exists \cite{Lew:projection_constant}
a projection $q_\lambda : \fcal_\lambda \to \ecal_\lambda$ such that
$\| q_\lambda \phi \|_\lambda \leq \sqrt{n} \| \phi \|_\lambda$, where $n = \dim \ecal$. 
We now choose two positive test functions, $f$ on the Lorentz group and $h$ on $\rbb_+$, both of compact support
and normalized in the respective $L_1$ norm,
and define $p_\lambda$ as 
\begin{multline}
   \sigma (p_\lambda \phi) :=
  \convol_{h} \Big( \mu \mapsto \int d\nu(\Lambda) \, f(\Lambda) \,  \sigma( \alpha_\Lambda q_{\mu} \alpha_\Lambda^{-1} \phi) \Big)(\lambda)
\\ \text{for } \sigma \in \cinftys, \; \phi \in \cinftyss.
\end{multline}
Here $\nu$ is the Haar measure on the Lorentz group, and $\convol_h$ 
is the map established in Lemma~\ref{lem:meanconvolution}. 
Since $\ecal$ is invariant under $\alpha_\Lambda$,
each $\alpha_\Lambda q_{\mu} \alpha_\Lambda^{-1}$ is a projector onto $\ecal$, and one easily sees 
in matrix elements that then the same is true for $p_\lambda$.
Using Lemma~\ref{lem:meanconvolution}~\ref{convolNorm}, we find the following estimate for $p_\lambda$:
\begin{equation} \label{pLambdaBound1}
|\sigma (p_\lambda \phi)| \leq 
 2 \log a\; \|h\|_\infty \|f\|_1 \sup_{\mu\in [1/a,a]} \sup_{\Lambda \in \supp f}
|\sigma( \alpha_\Lambda q_{\lambda\mu} \alpha_\Lambda^{-1} \phi)|.
\end{equation}
Here we can further estimate:
\begin{multline}  \label{pLambdaBound2}
|\sigma( \alpha_\Lambda q_{\lambda\mu} \alpha_\Lambda^{-1} \phi)|
\leq \sqrt{n}
\|\sigma(\uR_\lambda^{-\ell} \cdotarg \uR_\lambda^{-\ell} ) \|\;
\|\uR_\lambda^{\ell} \phi \uR_\lambda^{\ell}\|
\\
\times 
\| \uR_\lambda^{\ell}\alpha_\Lambda(\uR_\lambda^{-\ell}) \|^2
\|\uR_\lambda^{-\ell} \uR_{\mu\lambda}^{\ell} \|^2
\|\uR_\lambda^{\ell} \uR_{\mu\lambda}^{-\ell} \|^2
\| \uR_\lambda^{\ell} \alpha_\Lambda^{-1}(\uR_\lambda^{-\ell}) \|^2.
\end{multline}
Using spectral analysis of the $\uR_\lambda$, we can find a uniform
bound on the right-hand side when $\mu$ and $\Lambda$ range over a compact set.
Thus, combining Eqs.~\eqref{pLambdaBound1} and  \eqref{pLambdaBound2},
we obtain a constant $c$ (depending on $f$, $h$ and $\ell$) such that the
proposed estimate for $\lnorm{\up \; \uphi}{\ell}$ holds:
\begin{equation} \label{pLambdaBound3}
|\sigma (p_\lambda \phi)| \leq 
 c \|\sigma(\uR_\lambda^{-\ell} \cdotarg \uR_\lambda^{-\ell} ) \| \; \|\uR_\lambda^{\ell} \phi \uR_\lambda^{\ell}\|.
\end{equation}

It remains to show that the symmetry transforms act continuously on $\up \, \uphi$. 
Here continuity for the translations is clear; we check continuity of $\ualpha_{\mu,\Lambda} \up \, \uphi$
as $(\mu,\Lambda) \to \id$. Using the translation invariance of the Haar measure
and of the convolution $\convol_h$, we can derive the following for any $\lambda > 0$ and $\sigma \in \cinftyss$:
\begin{equation} \label{bigSymmSplit}
  \begin{aligned}
     \sigma( &(\ualpha_{\mu,\Lambda} \up \, \uphi - \up \, \uphi)_\lambda ) = \\
        & \convol[h(\mu^{-1} \cdotarg)-h] \Big( \mu' \mapsto \int d\nu(\Lambda') \, f(\Lambda') \,  
        \sigma( \alpha_{\Lambda\Lambda'} q_{\mu'} \alpha_{\Lambda'}^{-1} \uphi_{\mu\lambda}) \Big) (\lambda)\\
       +\,& \convol[h] \Big( \mu' \mapsto \int d\nu(\Lambda') \, f(\Lambda') \,  
        \sigma( \alpha_{\Lambda\Lambda'} q_{\mu'} \alpha_{\Lambda'}^{-1} (\uphi_{\mu\lambda}-\uphi_{\lambda})) \Big) (\lambda) \\
       +\,& \convol[h] \Big( \mu' \mapsto \int d\nu(\Lambda') \, \Big(f(\Lambda^{-1}\Lambda') - f(\Lambda') \Big) \,  
        \sigma( \alpha_{\Lambda'} q_{\mu'} \alpha_{\Lambda'}^{-1} \alpha_\Lambda \uphi_{\lambda}) \Big) (\lambda)\\
       +\,& \convol[h] \Big( \mu' \mapsto \int d\nu(\Lambda') \,  f(\Lambda') \,  
        \sigma( \alpha_{\Lambda'} q_{\mu'} \alpha_{\Lambda'}^{-1} (\alpha_\Lambda \uphi_{\lambda}-\uphi_{\lambda})) \Big)(\lambda)
  \end{aligned}
\end{equation}
Now we can use the following uniform estimates: Since $\uphi \in \uPhi$, we have $\lnorm{ \ualpha_\Lambda \uphi - \uphi}{\ell'} \to 0$ and $\lnorm{ \ualpha_\mu \uphi - \uphi}{\ell'} \to 0$ as $(\mu,\Lambda) \to \id$, where $\ell'$ is sufficiently large.
Further, since $g$ and $h$ are test functions, one has $\|g(\Lambda^{-1}\cdotarg) - g\|_1 \to 0$ and
$\|h(\mu^{-1}\cdotarg) - h\|_\infty \to 0$ in that limit. Applying all these to Eq.~\eqref{bigSymmSplit},
and using similar techniques as in Eqs.~\eqref{pLambdaBound1}--\eqref{pLambdaBound3}, we can obtain $\lnorm{\ualpha_{\mu,\Lambda} \up \, \uphi - \up \, \uphi}{\ell'} \to 0$. 
So $\up\,\uphi \in \uPhi$. 
\cmpqed\end{proof}

\section{Pointlike fields} \label{fieldSec}

In this section, our task will be to analyze the behavior of pointlike quantum fields in the scaling limit.
In order to relate these to the local algebras in question, we use the methods of 
\cite{Bos:short_distance_structure}. These methods are based on the assumption of 
a certain regularity condition, the \emph{microscopic phase space condition,} which we shall
recall in a moment. They allow for a full description of the field content of the net $\afk$
in the sense of Fredenhagen and Hertel \cite{FreHer:pointlike_fields}. 

We will introduce a phase space condition that is slightly stronger than
the one proposed in \cite{Bos:short_distance_structure}. Assuming this condition,
we show that the limit theory for pure limit states fulfills the original condition of \cite{Bos:short_distance_structure}.
We describe the scaling limit of pointlike fields in detail. In particular, 
we are able to recover the usual picture of multiplicative renormalization
of pointlike fields in our context.

\subsection{Phase space conditions}
 
Let us first recall the microscopic phase space condition from \cite{Bos:short_distance_structure}.
It demands that the natural inclusion map $\Xi: \cinftys \hookrightarrow \Sigma$ can be approximated by finite-rank maps, when the image functionals $\Xi(\sigma) \in \Sigma$ are restricted to small local algebras $\afk(r)$, $r \to 0$. The approximation quality, measured in the norms and seminorms introduced in Sec.~\ref{energyBoundsSec}, can be chosen to any given polynomial order in $r$.
The precise definition is as follows.

\begin{psc}\label{phase1} 
For every $\gamma \geq 0$ there exist an $\ell \geq 0$ and a map $\psi: \ccal^\infty(\Sigma) \to \Sigma$ of finite rank, such that
\begin{align*}
  \lnorm{\psi}{\ell} &<\infty, \\
  \lnorm{(\Xi-\psi) \restrict \afk(r) }{\ell} &= o(r^\gamma)\text{ as }r \to 0.
\end{align*}
\end{psc}

We shall call the map $\psi$ appearing above an \emph{approximating map (\ref{phase1}) of order $\gamma$},
with the roman numeral referring to the phase space condition. It is known that the image of the dual map $\psi\st$
essentially consists of pointlike fields. More precisely, let us consider for $\gamma \geq 0$ the following space:\footnote{
Note that this definition differs slightly from the convention chosen in \cite{Bos:short_distance_structure}. 
The effect of this change is that the map $\gamma \mapsto \dim \Phi_\gamma$ is guaranteed to be continuous from the right.}
\begin{equation}
  \Phi_\gamma = \bigsetprop{ \phi \in \cinftyss }{ \sigma(\phi) = 0 \text{ whenever } \|\sigma \restrict \afk(r)\| 
= O(r^{\gamma+\epsilon}) \text{ for some } \epsilon > 0}.
\end{equation}
We know from \cite{Bos:short_distance_structure} that, if $\psi$ is an approximating map (\ref{phase1}) 
of order $\gamma+\epsilon$ for some $\epsilon>0$, then $\Phi_\gamma \subset \img \psi\st$. If $\psi$ is of minimal
rank with this property, then equality holds.

As shown in \cite{Bos:short_distance_structure}, the phase space condition guarantees that 
the finite-dimensional spaces $\Phi_\gamma$ consist of pointlike quantum fields, which fulfill the
Wightman axioms after smearing with test functions. Their union $\PhiFH = \cup_\gamma \Phi_\gamma$
is equal to the field content of the theory as introduced by Fredenhagen and Hertel \cite{FreHer:pointlike_fields}.
The spaces are invariant under Lorentz transforms and other symmetries of the theory. 
Further, an operator product expansion exists between those fields \cite{Bos:product_expansions}.

However, the above Condition \ref{phase1} does not seem strict enough to guarantee a regular scaling limit
of the fields. 
This is, roughly, because
the estimates are not preserved under scaling: The short distance dimension $\gamma$ and the energy dimension
$\ell$ of the fields are not required to coincide. We therefore propose a stricter condition.

\begin{psc}\label{phase2} 
For every $\gamma \geq 0$ there exist $c, \epsilon, r_1>0$ and a map $\psi: \ccal^\infty(\Sigma) \to \Sigma$ of finite rank, such that
\begin{align*}
  \norm{\psi\restrict\Sigma(E),\afk(r)} &\leq c(1+Er)^\gamma, \quad &\text{for }E\geq 1, r\leq r_1,\\
   \norm{(\Xi-\psi) \restrict \Sigma(E),\afk(r) } & \leq c(Er)^{\gamma+\epsilon}  &\text{ for }E\geq 1, Er \leq r_1.
\end{align*}
Here the restriction $\restrict \Sigma(E), \afk(r)$ is to be understood as follows: the map (e.g. $\psi$) is restricted to $\Sigma(E)$, and its image points, being 
linear forms on $\afk$, are then restricted to $\afk(r)$. 
\end{psc}

Again, we shall call the map $\psi$ an \emph{approximating map (\ref{phase2}) of order $\gamma$}.
This stricter criterion, which is still fulfilled in free field theory in physical space-time \cite{Bos:operatorprodukte},
will allow us to pass to the scaling limit; only the scale-invariant expression $Er$ enters in its estimates.
For consistency, we show that the image of $\psi\st$ has similar properties to those implied by
Condition \ref{phase1}.

\begin{proposition}
If $\psi$ is an approximating map (\ref{phase2}) of order $\gamma$, then $\Phi_\gamma \subset \img \psi\st$.
If $\psi$ is of minimal rank with this property, then $\Phi_\gamma = \img \psi\st$.
\end{proposition}
\begin{proof}
For the first part, it suffices to show that $\psi(\sigma) = 0$ implies $\sigma \restrict \Phi_\gamma = 0$.
So let $\sigma \in \cinftys$ with $\psi(\sigma) = 0$. 
For any $E \geq 1$, set $\sigma_{E} = \sigma(P(E) \cdot P(E))$. 
Then, we can obtain for any $\ell>0$,
\begin{equation} \label{sigmaEEst}
  \|\sigma-\sigma_E\| \leq 2 \lnorm{\sigma}{\ell} (1+E)^{-\ell};
 \quad
  \lnorm{\sigma-\sigma_E}{\ell} \leq 2 \lnorm{\sigma}{2 \ell} (1+E)^{-\ell}.
\end{equation}
Let us consider the estimate
\begin{equation} \label{psiSigmaEst}
   \|\sigma \restrict \afk(r)\| = \| (\sigma-\psi(\sigma)) \restrict \afk(r)\|
   \leq \|(\Xi-\psi)(\sigma_E) \restrict\afk(r) \| + \|\sigma-\sigma_E\| + \|\psi(\sigma-\sigma_E)\|.
\end{equation}
Here we choose $E=r^{-\eta}$ with sufficiently small $\eta>0$, 
and observe that $\|\psi\|^{(\ell)} <\infty$ for large $\ell$, which is easily seen by expanding $\psi$ in a basis and applying Lemma~\ref{polyBoundsLemm}.
Using this, Eq.~\eqref{sigmaEEst}, and the fact that $\psi$ is approximating (\ref{phase2}), 
we can achieve that the right-hand side of \eqref{psiSigmaEst} vanishes
like $O(r^{\gamma+\epsilon'})$ for some $\epsilon'>0$. 
But that implies $\sigma\restrict \Phi_\gamma = 0$.  Hence $\Phi_\gamma \subset \img \psi\st$.

Now suppose that there is $\phi \in \img \psi\st$ with $\phi \not \in \Phi_\gamma$. Then there exists
$\sigma \in \cinftys$ with $\sigma(\phi) = 1$, $\|\sigma \restrict \afk(r)\| = O(r^{\gamma+\epsilon'})$
for some $\epsilon'>0$. We note the following: With $E \geq 1$ and $\ell > 0$ to be specified later, we have
per Eq.~\eqref{sigmaEEst},
\begin{multline}
  \| \psi(\sigma) \restrict \afk(r) \| \leq 
\| (\psi-\Xi)(\sigma_E) \restrict \afk(r) \| + \| (\psi-\Xi)(\sigma-\sigma_E ) \| + \|\sigma\restrict\afk(r)\| 
\\
\leq
  O((Er)^{\gamma+\epsilon})
  + 2 \lnorm{\psi-\Xi}{\ell} (1+ E)^{-\ell} \lnorm{\sigma}{2\ell} + O(r^{\gamma+\epsilon'}). 
\end{multline}
Choosing $E = r^{-\epsilon/2(\gamma+\epsilon)}$, and $\ell$ sufficiently large, we can certainly find
$\epsilon''>0$ such that $\| \psi(\sigma) \restrict \afk(r) \| = O(r^{\gamma+\epsilon''})$.
Also, it is clear that $\|\phi\restrict \Sigma(E)\| = O(E^\gamma)$.

Now consider the map $\hat \psi := \psi - \psi(\sigma) \phi$. We show that it is
also an approximating map (\ref{phase2}) of order $\gamma$, but of lower rank than $\psi$.
First we compute for $r \leq r_1$, $E \geq 1$:
\begin{multline}
  \| \hat \psi \restrict \Sigma(E), \afk(r)\| 
 \leq \| \psi \restrict \Sigma(E), \afk(r)\| + \| \psi(\sigma) \restrict \afk(r) \| \,\|\phi \restrict \Sigma(E)\|
\\
 \leq c (1+Er)^\gamma + O(r^{\gamma+\epsilon''}) O(E^\gamma) \leq c'(1+Er)^\gamma
\end{multline}
with some $c'>0$. This is the first of the desired estimates. The second estimate follows similarly, for $Er \leq r_1$:
\begin{multline}
  \| (\Xi-\hat \psi) \restrict \Sigma(E), \afk(r)\| 
 \leq \| (\Xi- \psi) \restrict \Sigma(E), \afk(r)\| + \| \psi(\sigma) \restrict \afk(r) \| \,\|\phi \restrict \Sigma(E)\|
\\
 \leq c (Er)^{\gamma+\epsilon} + O(r^{\gamma+\epsilon''}) O(E^\gamma) \leq c'(Er)^{\gamma+\epsilon''}.
\end{multline}
(We have assumed $\epsilon\geq \epsilon''$ here.)  So $\hat\psi$ is an approximating map (\ref{phase2}) of order $\gamma$.
Also, it is clear that $\img \hat \psi\st \subset \img \psi\st$, and thus $\ker \psi \subset \ker \hat\psi$.
We know that $\sigma \not \in \ker \psi$, since $\sigma(\phi) = 1$. On the other hand,
$\hat \psi(\sigma) = \psi(\sigma)-\psi(\sigma)\sigma(\phi) = 0$. Thus $\ker \hat\psi$ is strictly larger than
$\ker \psi$, implying $\rank \hat \psi < \rank \psi$. If $\rank \psi$ was minimal, this is not possible;
thus we must have $\img \psi\st = \Phi_\gamma$.
\cmpqed\end{proof}

We will formulate another phase space condition which will be of technical importance for us,
and generalizes the above inasmuch as the finite-rank maps $\psi$ are allowed to depend on $r$
in a controlled way.

\begin{psc}\label{phase3}For every fixed $\gamma \geq 0$ there exist $c,\epsilon,r_1>0$, a closed subspace 
$\kcal \subset \ccal^\infty(\Sigma)$ of finite codimension, and for each $r \leq r_1$ a map $\psi_r: \ccal^\infty(\Sigma) \to \afk(r)_*$, such that $\kcal \subset \ker \psi_r$ and
\begin{align*}
  \norm{\psi_r\restrict\Sigma(E)} &\leq c(1+Er)^\gamma, \quad &\text{for }E\geq 1, r\leq r_1,\\
   \norm{(\Xi\restrict\afk(r)-\psi_r) \restrict\Sigma(E) } & \leq c(Er)^{\gamma+\epsilon} 
&\text{for }E\geq 1, Er \leq r_1.
\end{align*}
\end{psc}

These maps $\psi_r$ will be called approximating maps (\ref{phase3}) of order
$\gamma$. We now show how these phase space conditions are interrelated. First, we deduce from Condition~\ref{phase3} a version
relating to the energy norms $\lnorm{\cdotarg}{\ell}$ rather than to a sharp cutoff.
\begin{lemma} \label{cutoffChangeLemm}
Let $\{\psi_r\}$ be a set of approximating maps (\ref{phase3}) of order $\gamma$.
Then, we can find $\ell > 0$ and $c'>0$ such that
\begin{align*}
  \lnorm{\psi_r}{\ell} &\leq c' & \text{for all } r\leq r_1,\\
  r^{-\gamma} \lnorm{(\Xi\restrict\afk(r)-\psi_r) }{\ell} &\to 0 &\text{ as }r \to 0.
\end{align*}
\end{lemma}
\begin{proof}
The first part follows by expressing $\psi_r$ in a basis and applying 
Lemma~\ref{polyBoundsLemm}.---For the second part, set $\varphi_r = \Xi\restrict\afk(r) - \psi_r$, 
and let $\ell,\ell'>0$ be sufficiently large in the following. 
In the expression
\begin{equation}
  \lnorm{\varphi_r}{\ell+\ell'} =
      \|  \varphi_r( \idop \cdot R^{\ell+\ell'} \cdotarg R^{\ell+\ell'} \cdot \idop )  \| \, ,
\end{equation}
we replace the identities shown as $\idop$ with $(\idop-P(E)) + P(E)$.
A brief calculation shows that
\begin{equation} \label{ellPlusEllP}
  \lnorm{\varphi_r}{\ell+\ell'} \leq
    2 \lnorm{\varphi_r}{\ell} (1+E)^{-\ell'} + \| \varphi_r \restrict \Sigma(E) \| 
%\\
  \leq 2c' (1+E)^{-\ell'}   + c (Er)^{\gamma + \epsilon} 
\end{equation}
for $E \geq 1$, $Er \leq r_1$. Now setting $E = r^{-\eta}$
with sufficiently small positive $\eta$, and choosing $\ell'$ large
enough, it is obvious that
\begin{equation}
  r^{-\gamma} \lnorm{\varphi_r}{\ell+\ell'}  \xrightarrow[r \to 0]{} 0;
\end{equation}
which gives the desired estimate after a redefinition of $\ell$.
\cmpqed\end{proof}

The relations between the different conditions are now:
\begin{proposition} \label{phaseCompareProp}
The following implications hold between the phase space conditions: 
\ref{phase2} $\implies$ \ref{phase3} $\implies$ \ref{phase1}.
\end{proposition}

\begin{proof}
The implication \ref{phase2} $\implies$ \ref{phase3} follows at once by defining $\psi_r := \psi\restrict\afk(r)$ and $\kcal := \ker \psi$. We prove \ref{phase3} $\implies$ \ref{phase1}, setting out from the estimates in Lemma~\ref{cutoffChangeLemm}.
Let $\gamma \geq 0$ be given, and consider the corresponding $\kcal$ and $\psi_r$. Define
\begin{equation}
\kcal_\perp := \{\phi\in\ccal^\infty(\Sigma)^*\,|\,\phi\restrict \kcal = 0\}.
\end{equation}
Since $\kcal$ is of finite codimension, $\kcal_\perp$ is finite dimensional. Furthermore $\kcal \subset \ker \psi_r$ implies $\img \psi_r^* \subset \kcal_\perp$ for all $r \leq r_1$. Now let $p = \sum_j\sigma_j\phi_j$ be a projector onto $\kcal_\perp$, i.e.\ a linear map $p : \ccal^\infty(\Sigma)^* \to \ccal^\infty(\Sigma)^*$ such that $p^2 = p$ and $\img p = \kcal_\perp$; and 
let $p_*:\ccal^\infty(\Sigma)\to\ccal^\infty(\Sigma)$ be its predual map,
which always exists since $\rank p$ is finite. 
It is easily seen that $p_*(\sigma)-\sigma \in \kcal$ for all $\sigma \in \ccal^\infty(\Sigma)$, so that $\psi_r\circ p_* = \psi_r$ for all $r\leq r_1$. Furthermore if $\ell'>0$ is so big that $\lnorm{\phi_j}{\ell'} < \infty$ for all $j$, it is clear that, for each $\ell > 0$,
\begin{align}
\lnorm{p_*}{\ell'} &= \sup_{\sigma \in \ccal^\infty(\Sigma)}\frac{\norm{p_*(\sigma)}}{\lnorm{\sigma}{\ell'}} \leq \sum_j \lnorm{\phi_j}{\ell'}\norm{\sigma_j}, \\
\norm{p_*}^{(\ell',\ell)} &= \sup_{\sigma \in \ccal^\infty(\Sigma)}\frac{\lnorm{p_*(\sigma)}{\ell}}{\lnorm{\sigma}{\ell'}} \leq \sum_j \lnorm{\phi_j}{\ell'}\lnorm{\sigma_j}{\ell}.
\end{align}
Therefore if we define $\psi := \Xi\circ p_*$ we have $\lnorm{\psi}{\ell'} = \lnorm{p_*}{\ell'}<\infty$, and if $\ell'\geq \ell$,
\begin{equation}\begin{split}
\lnorm{(\psi-\Xi)\restrict\afk(r)}{\ell'} &\leq \lnorm{\Xi\circ p_*\restrict\afk(r) - \psi_r}{\ell'}+\lnorm{\psi_r-\Xi\restrict\afk(r)}{\ell'}\\
&\leq \lnorm{(\Xi\restrict\afk(r)-\psi_r)\circ p_*}{\ell'} + \lnorm{\psi_r-\Xi\restrict\afk(r)}{\ell'}\\
&\leq \lnorm{\Xi\restrict\afk(r)-\psi_r}{\ell}\norm{p_*}^{(\ell',\ell)} + \lnorm{\psi_r-\Xi\restrict\afk(r)}{\ell} = o(r^\gamma),
\end{split}\end{equation}
which implies Condition~\ref{phase1}.
\cmpqed\end{proof}

\subsection{Phase space behavior of the limit theory} \label{phaseLimSec}

We will now investigate the phase space properties of the scaling limit theory. In what follows,
we shall assume that the original theory $\afk$ fulfills Phase space condition~\ref{phase2}, as introduced above. 
The goal of this section is to show that, if the mean $\mean$ is multiplicative, the limit theory $\afk_0$
fulfills at least the somewhat weaker Condition~\ref{phase1}, which however still allows for a full description
of pointlike fields.

We will keep $\gamma > 0$ and  $\psi$, an approximating map (\ref{phase2}) of order $\gamma$, fixed for this section. 
Our task is to
construct a map $\psi_0:\cinfty(\Sigma_0) \to \Sigma_0$ which fulfills the properties required in Condition~\ref{phase1}. Heuristically, one would like to define $\psi_0$ as a scaling limit
of the map $\psi$, such that
\begin{equation} \label{psiHeurDef}
  \psi_0( \pi_0\st \usigma) (\pi_0 \uA) = \lim_{\lambda \to 0} \psi(\usigma_\lambda) (\uA_\lambda). 
\end{equation}
Of course, this limit does not exist in general, and we need to replace it with the mean $\mean$. 
But even then, trying to use Eq.~\eqref{psiHeurDef} as a definition of $\psi_0$, it is not clear
why this would be well-defined in $\usigma$ and $\uA$. Specifically, it is not clear whether the right-hand side,
considered as a functional in $\uA$, is in the folium of the representation $\pi_0$.

For technical reasons, we will in fact use the dual map $\psi\st:\boundedops \to \cinftyss$ 
to define our phase space map in the limit, using the techniques developed in Sec.~\ref{energyBoundsSec}.
For $\uA \in \uafk(r_1)$, consider $\psi\st(\uA)$, i.e. the function $\lambda \mapsto \psi\st(\uA_\lambda)$. 
With a proper choice of $\psi$, these functions are elements of our space $\uPhi$.

\begin{lemma} \label{psicontLemm}
  Let Phase space condition \ref{phase2} be fulfilled. For $\gamma \geq 0$, 
 we can choose an approximating map (\ref{phase2}) 
 $\psi$ of order $\gamma$ and of minimal rank such that, 
for every $\uA \in \uafk(r_1)$, one has
 $\psi\st(\uA) \in \uPhi$. Further,
 $\lnorm{\psi\st(\uA)}{\ell} \leq c \|\uA\|$ for suitable $\ell,c$.
\end{lemma}
\begin{proof}
Let $\hat\psi$ be any approximating map (\ref{phase2}) of order $\gamma$ and of minimal rank. 
Phase space condition \ref{phase2} tells us that for suitable $\hat c$ and $\hat r_1$,
\begin{equation} \label{psiPolyEst}
  \| \hat\psi\st(\uA_\lambda) \restrict \Sigma(E/\lambda)\| \leq \hat c (1+Er)^\gamma \|\uA\| \;\; \text{ for } \uA \in \uafk(r), \; r \leq \hat r_1,
   \; E \geq 1, \; 0<\lambda\leq 1.
\end{equation}
By application of Lemma~\ref{polyBoundsLemm} with respect to $\lambda H$, 
this implies $\lnorm{\hat\psi\st(\uA)}{\ell} \leq c' \|\uA\|$ for sufficiently large $c',\ell$.
Now let $h$ be a positive test function of compact support on the Lorentz group, with $\|h\|_1=1$, and consider the map $\psi$,
defined by
\begin{equation}
  \psi(\sigma) = \int d\nu(\Lambda) \, h(\Lambda) \, \alpha_\Lambda \hat\psi (\alpha_\Lambda^{-1}\sigma),
\end{equation}
where the weak integral is well defined thanks to the fact that the restriction of $\alpha_\Lambda$ to $\Phi_\gamma$ is a finite-dimensional representation, and therefore is continuous in $\Lambda$. Then $\psi$ is as well an approximating map (\ref{phase2}) of order $\gamma$, with suitable constants $c > \hat c$, $r_1 < \hat r_1$, as is easily seen.
It fulfills $\lnorm{\psi\st(\uA)}{\ell} \leq c'' \|\uA\|$ for sufficiently large $c''$.
Also, it has the same rank as $\hat\psi$,
since $\img \hat\psi\st = \Phi_\gamma$ is stable under Lorentz transforms. 
Now let $\uA \in \uafk(r_1)$. We prove that $g \mapsto \ualpha_g \psi\st(\uA)$ is continuous
in $\lnorm{\cdotarg}{\ell}$ for large $\ell$. For translations, this is trivially fulfilled,
and for dilations it follows from $\ualpha_\mu \psi\st (\uA) = \psi\st (\ualpha_\mu \uA)$
and from the continuity properties of $\uA$. So let $\Lambda$ be in the Lorentz group.
Similar to Eq.~\eqref{bigSymmSplit}, we obtain
\begin{equation} 
  \begin{aligned}
     (\ualpha_\Lambda \psi\st(\uA) - \psi\st(\uA))_\lambda = 
       &  \int d\nu(\Lambda') \, \big(h(\Lambda^{-1}\Lambda') - h(\Lambda') \big) \,  
         \alpha_{\Lambda'} \hat\psi\st \alpha_{\Lambda'}^{-1} \alpha_\Lambda \uA_{\lambda} \\
       +& \int d\nu(\Lambda') \,  h(\Lambda') \,  
         \alpha_{\Lambda'} \hat\psi\st \alpha_{\Lambda'}^{-1} (\ualpha_\Lambda \uA-\uA)_{\lambda}
  \end{aligned}
\end{equation}
in the sense of matrix elements. Since $\Lambda \mapsto \ualpha_\Lambda \uA$ is norm continuous,
$h$ is smooth, and $\hat\psi$ fulfills the bounds in Eq.~\eqref{psiPolyEst}, the above
expression vanishes uniformly in $\lambda$ in some norm $\lnorm{\cdotarg}{\ell}$
as $\Lambda\to\id$. So $\psi \st (\uA) \in \uPhi$,
and $\psi$ has all required properties.
\cmpqed\end{proof}

Using the map $\psi$ constructed in the above lemma, we now define our approximation map in the limit theory by
\begin{equation} 
 \psi_0\st: \uafk(r_1) \to \cinfty(\Sigma_0)\st, \quad \psi_0\st (\uA) := \pi_0 (\psi\st(\uA)).
\end{equation}
Proposition~\ref{phiLimitProp} and Lemma~\ref{psicontLemm} yield the estimate
\begin{equation} \label{psi0Norm}
  \lnorm{ \psi_0\st(\uA) } {2\ell+1} \leq c  \|\uA\| \quad \text{ for } \uA \in \uafk(r), \; r \leq r_1 .
\end{equation}
This estimate also shows that the predual map $\psi_0 : \cinfty(\Sigma_0) \to \uafk(r_1)\st$ exists,
fulfilling $\psi_0(\sigma_0)(\uA) = \sigma_0 (\psi_0\st(\uA))$.
Spelling this out explicitly for $\sigma_0 = \pi_0\st \usigma$, we obtain
\begin{equation} \label{psi0Limit}
   \psi_0(\pi_0\st\usigma)(\uA) = \mean( \psi(\usigma_\lambda)(\uA_\lambda) ),
\end{equation}
which resembles the heuristic formula in Eq.~\eqref{psiHeurDef}.

In addition to the estimate \eqref{psi0Norm}, we also obtain from Condition~\ref{phase2} 
and Proposition~\ref{phiLimitProp} that
for normalized $\uA \in \uafk(r)$, 
\begin{equation} \label{psi0StDiff}
  \| (\psi_0\st(\uA)-\pi_0(\uA)) \restrict \Sigma_0(E-0) \| 
\leq
  c \|\uA\| (Er)^{\gamma + \epsilon},
\end{equation}
supposing that $E \geq 1$, $Er \leq r_1$. Since the right-hand side is continuous in $E$, 
this amounts to  
\begin{equation} \label{psi0Diff}
  \| (\psi_0-\pi_0\st\Xi_0) \restrict \Sigma_0(E),\uafk(r) \| 
\leq
   c (Er)^{\gamma + \epsilon}.
\end{equation}

So most parts of Phase space condition \ref{phase1} are fulfilled in the limit theory.
However, a crucial point needs to be investigated: whether $\psi_0$ is of finite
rank. We will actually show this in the case of our \emph{pure} limit states.

\begin{proposition} \label{limitRankProp}
%\todo{Prove more general version: For $\mean$ arbitrary, $\img \psi_0$ is a finite-dimensional
%module over $\pi_0(\zfk(\uafk))$, with dimension less or equal to $\rank\psi$. (Is it possible to obtain this?)}
If the mean $\mean$ is multiplicative, then $\rank \psi_0 \leq \rank \psi$.
\end{proposition}

\begin{proof}
Let $n := \rank \psi$. It suffices to prove the following:
For given $\uA_0,\ldots,\uA_n \in \uafk(r_1)$, there are constants
$(c_0,\ldots,c_n) \in \cbb^{n+1}\backslash \{0\}$ such that
\begin{equation} \label{lcLimit}
   \psi_0\st \Big(\sum_{j=0}^n c_j \uA_j\Big) = 0 ;
\end{equation}
for this shows that $\dim \img \psi_0 \leq n$. 

Let such $\uA_j$ be given.  Since $\psi$ is of rank $n$, it is certainly possible
to choose, for any $0<\lambda\leq1$, numbers $c_{0,\lambda},\ldots,c_{n,\lambda} \in \cbb$
such that
\begin{equation} \label{lcFiniteScale}
   \psi\st \Big(\sum_{j=0}^n c_{j,\lambda} \uA_{j,\lambda} \Big) = 0.
\end{equation}
Here not all of the $c_{j,\lambda}$ vanish, and so we can choose them to be on the
unit sphere in $\cbb^{n+1}$. Then the functions $\lambda \mapsto c_{j,\lambda}$
are bounded, and we can define $c_j:=\mean(c_{j,\lambda})$. 
We observe that for any $\usigma\in\uSigma$,
\begin{equation} \label{diffInKernel}
  \pi_0\st\usigma( \psi_0\st\big(\sum_j c_j \uA_j \big ))
 = \sum_j \mean( c_{j,\lambda}) \mean( \psi(\usigma_\lambda)(\uA_{j,\lambda}) )
 = \mean( \psi(\usigma_\lambda)\big(\sum_j c_{j,\lambda}\uA_{j,\lambda} \big) )  
 = 0,
\end{equation}
where we have used multiplicativity of the mean. 
Since we can extend this equation from $\pi_0\st\uSigma$ to $\cinfty(\Sigma_0)$,
this establishes Eq.~\eqref{lcLimit}. As a last point, not all of the $c_j$ vanish,
since $\sum_j \bar c_j c_j = \sum_j \mean(\bar c_{j,\lambda} c_{j,\lambda})=1$, again 
using the multiplicative mean. This proves $\rank \psi_0 \leq n$.
\cmpqed\end{proof}

We note that the same will in general \emph{not} be true if the mean is not multiplicative,
for example for invariant means. In this case, the image of $\psi_0\st$ will in general
be infinite-dimensional, containing in particular all operators in $\pi_0(\zfk(\uafk))$.

The remaining problem for establishing Phase space condition~\ref{phase1} is now the target space
of $\psi_0$: Its image points are not normal functionals with respect to $\pi_0$. 
This does not directly affect our computations here, but is crucial in the analysis of 
associated Wightman fields \cite{Bos:short_distance_structure}. We solve this problem by taking the normal
part of those functionals with respect to $\pi_0$, and showing that this is just as well suited for our
approximation. The notion of the normal part of a functional on a $C\st$ algebra with respect to a 
specific state needs explanation, since we use it in a slightly nonstandard way; 
it is treated in detail in Appendix~\ref{normalPartApp}.
Here we note only that the normal part depends both on the state and the algebra, and is not compatible
with restriction to subalgebras. We obtain from Theorem~\ref{normalpartThm} for each $r\leq r_1$ a linear map
$\normalpart[\uafk(r),\uomega_0]: \uafk(r)\st \to \afk_0(r)_\ast$ of norm 1.
Now we define
for each $r \leq r_1$ a map $\psi_{0,r}: \cinfty(\Sigma_0) \to \afk_0(r)_\ast$ by  
\begin{equation}
    \psi_{0,r} := \normalpart[\uafk(r),\uomega_0] \circ \rho_r \circ \psi_0,
\end{equation}
where $\rho_r$ is the restriction map $\uafk(r_1)\st \to \uafk(r)\st$.

\begin{proposition}
The maps $\psi_{0,r}$ fulfill the two estimates requested in Phase space condition~\ref{phase3}.
Moreover, if $\mean$ is multiplicative, we have $\rank \psi_{0,r} \leq \rank \psi$.
\end{proposition}

\begin{proof}
The first estimate of Condition~\ref{phase3}
follows from the corresponding estimate for $\psi_0$, see Eq.~\eqref{psi0Norm},
and the fact that $\|\normalpart[\uafk(r),\uomega_0]\| = 1$  by Theorem~\ref{normalpartThm} \ref{normalpartNormItem}.
For the second estimate,
we remark that, with $\rho_{0,r}$ being the restriction to $\afk_0(r)$,
\begin{multline}
\psi_{0,r} - \rho_{0,r}\Xi_{0} 
= \normalpart[\uafk(r),\uomega_0] \rho_r  \psi_0 - 
\normalpart[\uafk(r),\uomega_0] \pi_0\st \rho_{0,r}\Xi_0
\\
= \normalpart[\uafk(r),\uomega_0] \rho_r \big( \psi_0 - \pi_0\st \Xi_0 \big),
\end{multline}
cf. Theorem~\ref{normalpartThm} \ref{normalpartIdItem}.
We can now prove the second estimate of Condition~\ref{phase3} from Eq.~\eqref{psi0Diff}.
By expressing $\psi_0$ in a basis, it is also clear that composition with 
$\normalpart[\uafk(r),\uomega_0]$ does not increase the rank of the map; 
hence $\rank \psi_{0,r} \leq \rank \psi$ for multiplicative $\mean$ 
by Proposition~\ref{limitRankProp}.
\cmpqed\end{proof}

Setting $\kcal := \ker \psi_0 \subset \ker \psi_{0,r}$, the limit theory then fulfills
Phase space condition~\ref{phase3} by virtue of the maps $\psi_{0,r}$.
Due to Proposition~\ref{phaseCompareProp},
this implies Phase space condition~\ref{phase1} for the limit theory.
The dimensions of the field spaces $\Phi_\gamma$ do not increase when passing to the limit,
since by our construction, the field space $\Phi_{0,\gamma}$ of the limit theory
is contained in $\img \psi_0\st$.

\begin{theorem} \label{limitApscThm}
If the original net $\afk$ fulfills 
Phase space condition~\ref{phase2}, 
and the mean $\mean$ is multiplicative, then the scaling limit net 
$\afk_0$ fulfills Phase space condition~\ref{phase1}.
For the size of the field content, we have
\begin{equation*}
   \dim \Phi_{0,\gamma} \leq \dim \Phi_\gamma.
\end{equation*}
\end{theorem}

This establishes all the consequences of the phase space condition in the limit theory,
including the existence of operator product expansions. We shall see this more explicitly in Sec.~\ref{opeSec}.

\subsection{Renormalized pointlike fields} \label{renormSec}

We can now describe the renormalization limit of pointlike fields in our context. Heuristically,
as noted in the introduction, renormalized point fields should appear
as images of operator sequences $\uA \in \afk$, which already bear the ``correct'' renormalization, 
under the finite-rank map $\psi\st$. We shall now investigate this in detail.

In the following, let $\gamma > 0$ and a corresponding approximating map (\ref{phase2}) $\psi$ 
of order $\gamma$ be fixed,
where we choose $\psi$ as described in Lemma~\ref{psicontLemm}. 
For $\uA \in \uafk(r)$, $r \leq r_1$, also kept fixed for the moment, we set $\uphi := \psi\st(\uA)$.
Lemma~\ref{psicontLemm} guarantees that $\uphi\in\uPhi$, i.e.~$\uphi$ is a ``correctly renormalized''
sequence. By Proposition~\ref{phiLimitProp}, we know
that a well-defined scaling limit $\pi_0(\uphi) \in \cinfty(\Sigma_0)$ exists in the limit theory.
A priori, we do not know anything about localization properties of $\pi_0(\uphi)$ however.
For describing these, we will establish a uniform approximation of $\uphi$ with bounded operators.

\begin{theorem} \label{fieldApproxThm}
Let $\uphi\in \uPhi$ such that
$\uphi_\lambda \in \PhiFH$ for all $\lambda$. 
There exist operators $\uA_r \in \uafk(r)$,
$0<r\leq 1$, and constants $k, \ell > 0$ such that in the limit $r \to 0$:
\begin{enumerate}
\localitemlabels

\item \label{polyNorm} \quad $\| \uA_r \| = O(r^{-k})$,
\item \label{uniApp} \quad $\lnorm{ \uA_r - \uphi} {\ell} = O(r)$,
\item \label{limitApp}\quad $\lnorm{ \pi_0(\uA_r) - \pi_0(\uphi)} {\ell} = O(r)$.

\end{enumerate}
\end{theorem}

\begin{proof}
We use methods similar to \cite[Lemma 3.5]{Bos:short_distance_structure}. Choose a positive test function
$f \in \scal(\rbb^{s+1})$ with $\|f\|_1 = 1$, and with support in the double cone $\ocal_{r=1/2}$.
Further, set $f_r(x) = r^{-(s+1)} f(x/r)$, which is then a ``delta sequence'' as $r \to 0$.
For any fixed $r$ and $\lambda$, we know that $(\ualpha[f_r] \uphi)_\lambda = \int dx f_r(x) (\ualpha_x\uphi)_\lambda$ is a closable operator \cite{FreHer:pointlike_fields};
let $V_{r,\lambda} D_{r,\lambda}$ be the polar decomposition of its closure, 
with $V_{r,\lambda}$ being a partial isometry,
and $D_{r,\lambda}\geq 0$. We know from \cite{FreHer:pointlike_fields} that both 
$V_{r,\lambda}$ and the spectral projectors of $D_{r,\lambda}$ are contained in $\afk(\lambda r/2)$.
Let $\ell$ be sufficiently large such that $\lnorm{\uphi}{\ell} < \infty$, and set for $\epsilon > 0$:
\begin{equation}
  \uB_{r,\epsilon,\lambda} = \epsilon^{-1} V_{r,\lambda} \sin(\epsilon D_{r,\lambda}) \in \afk(\lambda r/2).
\end{equation}
This $\uB_{r,\epsilon}$ is certainly an element of $\ubfk$, but not necessarily of
$\uafk$. Using the inequality for real numbers,
\begin{equation}
  (x - \epsilon^{-1} \sin(\epsilon x)) ^2  \leq \epsilon^2 x^4\quad
 \text{for } x \geq 0, \; \epsilon > 0,
\end{equation}
we can use corresponding operator inequalities to establish the following estimate:
\begin{equation} \label{AvsPhiEst}
\begin{split}
\|  (\uB_{r,\epsilon,\lambda} - (\ualpha[f_r] \uphi)_\lambda ) \uR_\lambda^{4 \ell}\|^2
&= \| (D_{r,\lambda} - \epsilon^{-1} \sin(\epsilon D_{r,\lambda}) ) \uR_\lambda^{4 \ell} \|^2
\\
&\leq \epsilon^2 \| (\ualpha[f_r] \uphi)_\lambda\st (\ualpha[f_r] \uphi)_\lambda \uR_\lambda^{4 \ell} \|^2.
\end{split}
\end{equation}
Now employing the commutation relation in Eq.~\eqref{phiRComm}, and using estimates of the type
$\lnorm{ \ualpha[f'] \uphi }{\ell} \leq \|f'\|_1 \cdot \text{const.}$, we can obtain the
following uniform estimate in $\lambda$:
\begin{equation}
 \| (\ualpha[f_r] \uphi)_\lambda\st (\ualpha[f_r] \uphi)_\lambda \uR_\lambda^{4 \ell} \| \leq c r^{-4\ell},
\end{equation}
where the constant $c$ depends on the details of the function $f$, but not on $r$ or $\lambda$.
Combined with Eq.~\eqref{AvsPhiEst}, this yields
\begin{equation}
  \lnorm{\uB_{r,\epsilon} - \ualpha[f_r] \uphi }{4 \ell} \leq c \,\epsilon \, r^{-4\ell}.
\end{equation}
Using the spectral properties of the translation group, it is also easy to verify that
\begin{equation}
  \lnorm{ \uphi- \ualpha[f_r] \uphi}{\ell+1} = O(r).
\end{equation}
Now setting $\epsilon = r^{4\ell+1}$, and then redefining $\ell$, we have obtained operators
$\uB_r \in \ubfk$, with $\uB_{r,\lambda} \in \afk(\lambda r/2)$, and $k,\ell>0$, such that
\begin{equation} \label{bEstimates}
  \|\uB_r\| = O(r^{-k}), \quad \lnorm{ \uB_r -\uphi}{\ell} = O(r).
\end{equation}

In general, however, we cannot show that $\uB_r \in \uafk(r)$. In order to remedy this problem, we proceed
in two steps by regularizing first with respect to Poincar\'e transformations and then with respect to
dilations.
To that end, choose a family $(h_q^P)_{q>0}$ of positive test functions on $\poincare$, with compact supporting shrinking
to the identity as $q \to 0$, and converging to the delta function at $(x,\Lambda)=\id$.
We then set $\uC_r := \ualpha[h^P_q] \uB_r = \int d\nu(x,\Lambda) h^P_q(x,\Lambda) \ualpha_{x,\Lambda} \uB_r$. If $q$ is sufficiently small for $r$, 
we obtain $\uC_{r,\lambda}  \in \afk(3\lambda r/4)$, and that $(x,\Lambda) \mapsto \ualpha_{x,\Lambda}(\uC_r)$
is norm continuous. Also, it is clear that
$\|\uC_r\| = O(r^{-k})$, regardless of our choice of $q(r)$. If $q$ is small enough and taking $\ell > 1$, we have, by
spectral analysis of $\alpha_\Lambda(\uR_\lambda)\uR_\lambda^{-1}$, 
\begin{equation}
\| \ualpha[h^P_q]\uB_r - \ualpha[h^P_q]\uphi\|^{(\ell)}\leq c' \|\uB_r-\uphi\|^{(\ell)}
\end{equation} 
for some constant $c'>0$. Taking into account the continuity in some $\ell$-norm of $\uphi$ under Poincar\'e transformations,
and choosing $q(r)$ small enough, we can obtain from Eq.~\eqref{bEstimates} that
\begin{equation}
 \lnorm{ \uC_r - \uphi }{\ell} = O(r).
\end{equation}
Let now $1<a<4/3$, and let $h_a^D$ be a positive, continuous function of compact support in $(1/a,a)$, with
$\int h_a^D(\mu) d\mu/\mu = 1$ and $\|h_a^D\|_\infty \log a \leq 2$.
(The value of $a$ will be specified later, dependent on $r$.) 
Recalling Lemma~\ref{normContLemm},
set $\uA_r := \udelta[h_a^D]\uC_r\in\uafk(r)$. Clearly $\| \uA_r\| = O(r^{-k})$. 
Further, define $\uphi_a \in \uPhi$ by 
\begin{equation}
  \sigma(\uphi_{a,\lambda}) = \convol[h_a^D] \Big( \mu \mapsto \sigma(\uphi_\mu) \Big)(\lambda) = \int \frac{d\mu}{\mu} h_a^D (\mu) \sigma(\uphi_{\mu\lambda}).
\end{equation}
(The equality with the integral follows from continuity properties of $\ualpha_\mu \uphi$, and it is 
easily checked that in fact $\uphi_a \in \uPhi$.) We now consider the estimate
\begin{equation} \label{aphiSplit}
   \lnorm{\uA_r-\uphi}{\ell} \leq \lnorm{\uA_r-\uphi_a}{\ell} + \lnorm{\uphi_a-\uphi}{\ell}.
\end{equation}
For the first term on the right hand side, we have
\begin{equation}
   \sigma(\uA_{r,\lambda}-\uphi_{a,\lambda}) = \convol[h_a^D] \Big(\mu \mapsto \sigma(\uC_{r,\mu} - \uphi_\mu ) \Big)(\lambda)
   \quad\text{for all } \sigma \in\cinftyss,
\end{equation}
which yields the estimate
\begin{equation}
   \lnorm{\uA_r-\uphi_a}{\ell} \leq  2 (\log a) \|h_a^D\|_\infty \lnorm{\uC_r-\uphi}{\ell} 
   \sup_{\substack{\lambda >0\\\mu\in(1/a,a)}}\|\uR_\lambda/\uR_{\mu\lambda}\|^{2\ell} = O(r),
\end{equation}
regardless of our choice of $a$. The second term in Eq.~\eqref{aphiSplit} can be written in
terms of integrals; that gives
\begin{equation}
   \lnorm{\uphi_a-\uphi}{\ell} \leq  \|h_a^D\|_1 \sup_{\mu\in(1/a,a)}  \lnorm{\ualpha_\mu\uphi-\uphi}{\ell}.
\end{equation}
Now if $a(r)$ is chosen sufficiently close to $1$, we can certainly achieve that this bound vanishes
like $O(r)$, due to the continuity of $\mu \mapsto \ualpha_\mu \uphi$.
Inserting into Eq.~\eqref{aphiSplit}, we obtain $\lnorm{\uA_r-\uphi}{\ell} = O(r)$ as proposed,
which establishes part \ref{uniApp} of the proposition. Part \ref{polyNorm}
was already clear. For part \ref{limitApp}, we only need to invoke Proposition~\ref{phiLimitProp}.
\cmpqed\end{proof}

The above theorem shows in particular that $\pi_0 \uphi$ can be approximated with
bounded operators in the limit theory, with their localization region shrinking to the origin.
Applying the results of \cite{FreHer:pointlike_fields}, we can state:

\begin{corollary} \label{limitFieldCor}
  Let $\uphi\in\uPhi$ such that $\uphi_\lambda \in \PhiFH$ for all $\lambda$. 
  Then $\pi_0\uphi$ is an element of $\PhiFHLim$, the field content of the limit theory $\afk_0$.
\end{corollary}

This applies in particular to $\uphi = \psi\st(\uA)$, so $\pi_0 \uphi = \psi_0(\uA) \in \PhiFHLim$.
This relation holds true even in the case where the rank of $\psi_0$ is not finite.
For multiplicative means $\mean$, we further know by our results in Sec.~\ref{phaseLimSec}
that every local field in the limit theory can be obtained in this way; ``no new fields appear
in the limit''.

We now explain how our results are related to the usual formalism of renormalized fields.
As above, we consider $\uphi = \psi\st(\uA)$ with 
a fixed $\uA \in \uafk(r_1)$. We write the finite-rank map $\psi$ in a basis,
$\psi = \sum_j \sigma_j \phi_j$, with local fields $\phi_j$ associated with the original theory $\afk$. 
Now we have
\begin{equation} \label{renormFactor}
  (\ualpha_x \uphi)_\lambda = \alpha_{\lambda x} \psi\st(\uA_\lambda) 
 = \sum_j \sigma_j(\uA_\lambda) \phi_j(\lambda x)
 = \sum_j Z_{j,\lambda} \phi_j(\lambda x),
\end{equation}
defining the ``renormalization factors'' $Z_{j,\lambda} := \sigma_j(\uA_\lambda)$. 
By our above results, $\phi_0 := \pi_0 \uphi = \pi_0 \psi\st(\uA)$ is a local field
in the limit theory,
for which we have the formula
\begin{equation}
  \phi_0(x) = \pi_0 \ualpha_{x} \uphi = 
\text{``} \lim_\lambda  \text{''} \sum_j Z_{j,\lambda} \phi_j(\lambda x).
\end{equation}
The ``limit'' on the right-hand side needs to be read as an application of the mean $\mean$ to the appropriate
expectation values, i.e.
\begin{multline}
  \mtxe{\pi_0(\uB)\Omega_0}{\phi_0(x)}{\pi_0(\uB')\Omega_0} = 
 \mean \Big( \sum_j Z_{j,\lambda} \mtxe{\uB_\lambda\Omega}{  \phi_j(\lambda x)}{\uB_\lambda' \Omega} \Big)
\\
  \text{for } \uB,\uB' \in \bigcup_E \tilde\uafk(E).
\end{multline}
Symmetry transformations are compatible with this limit by Eq.~\eqref{phiLimitSymm}, 
so the symmetry group at finite scales
converges to the symmetry group in the limit. In the case of an invariant mean $\mean$, 
not only the Poincaré transformations but also the dilations can be extended in this way to the limit theory,
and hence to the fields. We obtain
\begin{equation} \label{dilationLim}
  U_0(\mu) \phi_0 U_0(\mu)\st = \pi_0 \ualpha_{\mu} \uphi, 
\quad \text{where }
   (\ualpha_{\mu} \uphi)_\lambda = \uphi_{\mu\lambda} = \sum_j Z_{j,\mu\lambda} \phi_j.
\end{equation}
So a shifting of the renormalization factors corresponds to a unitary transformation
of the fields in the limit theory. We may interpret $\ualpha_\mu$ as an action
of the renormalization group on the theory, in the sense of
Gell-Mann and Low \cite{GelLow:rengroup}. Thus the renormalization group $\ualpha_\mu$
induces the dilation symmetry $\alpha_{0,\mu}$ in the scaling limit.
Note however that the field spaces will in general not be finite dimensional 
in the limit if the mean $\mean$ is not multiplicative; so the representation
$\mu \mapsto \mathrm{ad}\, U_0(\mu) \lceil \Phi_\gamma$ is not finite dimensional. 

To understand this in more detail in an example, let us consider the case where
the limit theory factorizes as a tensor product like in Eq.~\eqref{limitTens}, such as in the case of a free 
field \cite{BDM:scaling2}. Setting $\zfk_0 := \pi_0(\zfk(\uafk))''$, a commutative algebra, 
we would have
\begin{equation} \label{algebraFactorizing}
\afk_0(\ocal) \isom \zfk_0  \wtens \hat \afk(\ocal), \quad 
 U_0(g) \isom U_0(g)\restrict \hcal_\zfk \otimes \hat U(g),
\end{equation} 
where $\hat\afk$ is the theory associated with a \emph{pure} limit state. 
Here $U_0(g)\restrict \hcal_\zfk=\idop$ for all Poincaré transformations.
The field content of the limit theory is characterized \cite{FreHer:pointlike_fields}
by
\begin{multline}
  \phi_0 \in \PhiFHLim \equivalent
\\
  R_0^\ell \phi_0 R_0^\ell 
  \in \bigcap_\ocal \overline{ R_0^\ell \afk_0(\ocal)  R_0^\ell }
 \isom  \bigcap_\ocal \overline{ \zfk_0 \otimes \hat R^\ell \hat\afk(\ocal)  \hat R^\ell }
\quad\text{for some $\ell>0$}.
\end{multline}
The intersection runs over all neighborhoods $\ocal$ of the origin, and the bar denotes 
weak closure. $\PhiFHLim$ in particular includes the finitely
generated modules $\zfk_0 \hat\Phi_\gamma$, whre $\hat\Phi_\gamma$ are the field spaces corresponding
to the theory $\hat\afk$. For a pure limit state, we
have $\zfk_0 = \cbb\idop$, so that $\PhiFHLim$ coincides with $\PhiFHHat$,
the field content of $\hat\afk$.

Now choose a special type of element
in $\PhiFHLim$, of the form $\phi_0=\idop\otimes\hat\phi$ with $\hat\phi \in \PhiFHHat$. 
Suppose that $\uphi\in\uPhi$ exists with $\pi_0 \uphi = \phi_0$ (we note that this is 
actually the case for a free field). 
For this special choice, Eq.~\eqref{dilationLim} reads
\begin{equation} 
  U_0(\mu) (\idop \otimes \hat\phi) U_0(\mu)\st = \idop \otimes \hat U(\mu) \hat \phi \hat U(\mu)\st 
 = \text{``lim''} \sum_j Z_{j,\mu\lambda} \phi_j.
\end{equation}
Here the scaling tranformations actually act like $\hat U \restrict \hat\Phi_\gamma$,
a finite dimensional representation 
of the dilation group; these are well classified \cite[Ch. V, \S 9]{Boe:gruppen_en}.
For more general elements of $\PhiFHLim$, the nontrivial action of dilations on the center
has to be taken into account, as per Eqs.~\eqref{dilationLim} and \eqref{algebraFactorizing}.

\section{Operator product expansions} \label{opeSec}

A critical point in the analysis of the scaling limit
is the behavior of the interaction with the changing scales. 
In Lagrangian quantum field theory,
this is usually formulated in terms of the \emph{renormalization group flow:}
A change in scale is compensated by a change in the Lagrangian,
i.e.~by modifying the coupling constants of the theory.
Here, we do not assume that the theory under discussion is generated 
by a Lagrangian, and the concept of coupling constants is unavailable
in our model-independent context. Rather, a change of interaction 
shows up in the algebraic relations of the observables, which are different
at each scale $\lambda$. Relating to pointlike quantum fields,
their (singular) algebraic structure is described by the operator
product expansion \cite{WilZim:products}. It is the behavior of this 
expansion at small scales which reflects the ``structure constants'' 
of our ``improper algebra'' of quantum fields.

We know that, as a consequence of the phase space condition,
an operator product expansion (OPE) exists for the theory at finite scales
\cite{Bos:product_expansions}. In this section, we will investigate how this
carries over to the limit theory. We assume throughout this section
that the original theory $\afk$ fulfills Phase space condition \ref{phase2}.

We will focus here on the OPE for the product of two fields, 
understood in the sense of distributions.
Recall from \cite{Bos:product_expansions} that the OPE at finite scales is roughly given by
\begin{equation}
  \phi(f) \phi'(f') \approx p_{\gamma'} ( \phi(f)\phi'(f')),
\end{equation}
where $\phi,\phi' \in \Phi_\gamma$ are pointlike fields,
$\gamma'$ is large enough for $\gamma$, and $p_{\gamma'}$
is a projector onto $\Phi_{\gamma'}$. The approximation is valid
in the limit where the support of the test functions $f$ and $f'$
shrinks to the origin.

In the following, let  $\uA, \uA' \in \uafk(r_1)$ be fixed, 
as well as $\gamma>0$, and let $\psi$ be an approximating map of order $\gamma$,
as in Lemma~\ref{psicontLemm}. 
We set $\uphi:=\psi\st(\uA)$, $\uphi':=\psi\st(\uA')$. Further, let
$f,f' \in \scal(\mkr)$ be fixed test functions with support 
in $\ocal_{r=1}$. We consider for $d>0$
\begin{equation}
  f_d{}\evp := d^{-(s+1)} f\evp (x/d);
\end{equation}
our short distance limit will then be $d \to 0$. 
We wish to analyze the product
\begin{equation} \label{oprodDef}
  \oprod := (\ualpha[f_d] \uphi) \cdot (\ualpha[f_d'] \uphi') \in \uPhi.
\end{equation}
See Eq.~\eqref{smearedField} for the notation. 
We note that each $(\ualpha[f_d] \uphi)_\lambda$ can be extended \cite{FreHer:pointlike_fields}
to an unbounded operator on the invariant domain $\cinftyh = \cap_{\ell>0} R^\ell \hcal$,
so the product is well-defined; and energy bounds and continuity properties 
with respect to $\ualpha_g$ can be obtained using  Eq.~\eqref{phiRComm}, so that in fact $\oprod \in\uPhi$.

Our task is to obtain a product expansion for $\oprod$, uniform at all scales.
To that end, we choose $\uA_r$, $\uA_r'$ as approximating sequences 
for $\uphi$, $\uphi'$ by Theorem~\ref{fieldApproxThm}. We set
\begin{equation}
  \uB_r :=  (\ualpha[f_d] \uA_r) \cdot (\ualpha[f_d'] \uA_r') \in \uafk(r+d).
\end{equation}
This sequence is supposed to approximate $\oprod$ as $r \to 0$.
In fact, we show:

\begin{lemma} \label{prodApproxLemm}
There are constants $c,\ell > 0$ such that
\begin{equation*}
   \lnorm{\oprod-\uB_r}{\ell} \leq c \,r\, d^{-\ell} \quad \text{for all } d>0 \text{ and } r \leq 1.
\end{equation*}
\end{lemma}

\begin{proof}
We write the difference $\oprod-\uB_r$ in terms of the individual fields:
\begin{equation}
\oprod-\uB_r = \big( \ualpha[f_d](\uphi-\uA_r) \big) \big(\ualpha[f_d']\uphi'\big)
+ \big(\ualpha[f_d]\uA_r \big) \big( \ualpha[f_d'](\uphi'-\uA_r') \big) .
\end{equation}
We shall derive estimates only for the first summand, the second is treated in an 
analogous way. Using the commutation relation in Eq.~\eqref{phiRComm} multiple times,
we obtain a relation of the type
\begin{equation}\label{lshiftedSum}
\uR^{2\ell} \big( \ualpha[f_d](\uphi-\uA_r) \big) (\ualpha[f_d']\uphi') \uR^{2\ell} 
= \sum_j c_j \uR^{\ell_j} \big(\ualpha[\partial_0^{n_j}f_d](\uphi-\uA_r)\big)
  \uR^{\ell_j'} \big(  \ualpha[\partial_0^{n_j'}f_d'](\uphi') \big) \uR^{\ell_j''},
\end{equation}
with certain constants $c_j$, where we can achieve that $\ell_j \geq \ell$, $\ell_j' \geq 2 \ell$, $\ell_j'' \geq \ell$,
$n_j \leq \ell$, $n_j' \leq \ell$. If now $\ell$ is sufficiently large
(as in Theorem~\ref{fieldApproxThm}), we can apply the following estimates.
\begin{alignat}{2}
\lnorm{\ualpha[\partial_0^{n_j}f_d](\uphi-\uA_r)}{\ell} &\leq
\|\partial_0^{n_j}f_d\|_1 O(r) &\leq d^{-\ell} O(r),
\\
\lnorm{\ualpha[\partial_0^{n_j'}f_d'](\uphi')}{\ell} &\leq
\|\partial_0^{n_j'}f_d'\|_1 O(1) &\leq d^{-\ell} O(1).
\end{alignat}
Applying this to Eq.~\eqref{lshiftedSum}, and using a similar bound
for $\uphi'$ and $\uphi$ exchanged, yields the proposed estimates after a redefinition of $\ell$.
\cmpqed\end{proof}

Now let $\ell'>\ell$ and $\gamma'>0$ be fixed; their value will be specified later.
We choose a uniform projector $\up$ onto
$\Phi_{\gamma'}$ according to Proposition~\ref{uniformProjProp} (for details see there),
where $\lnorm{\up\,\uphi}{\ell'} \leq \lnorm{\uphi}{\ell'}\cdot \text{const}$. We note
that $\up$ may depend on $\ell'$ and $\gamma'$, but the said estimate does not,
apart from a multiplicative constant. By Lemma~\ref{prodApproxLemm} above, we know
\begin{equation} \label{prodApprox}
   \lnorm{\oprod - \uB_r}{\ell'} = d^{-\ell} O(r),
\quad
   \lnorm{\up(\oprod - \uB_r)}{\ell'} = d^{-\ell} O(r).
\end{equation}
We now choose an approximating map (\ref{phase2}) $\psi'$ of order $\gamma'$. Then
$\up \psi'{}\st = \psi'{}\st$, and therefore  
\begin{multline}
   \lnorm{\uB_r-\up\uB_r}{\ell'}
\leq \lnorm{\uB_r-\psi'{}\st(\uB_r)}{\ell'} + \lnorm{\up(\uB_r-\psi'{}\st(\uB_r))}{\ell'}
\\
\leq \lnorm{\uB_r-\psi'{}\st(\uB_r)}{\ell'} \cdot \text{const}.
\end{multline}
With similar arguments as in Eq.~\eqref{ellPlusEllP}, we can obtain 
an estimate of the form
\begin{equation}
   \lnorm{\uB_r-\psi'{}\st (\uB_r)}{\ell'}
\leq \|\uB_r\| \big( (E(r+d))^{\gamma'} + (1+E)^{-\ell'/2} \big) \cdot \text{const},
\end{equation}
where we can choose $E$ dependent on $d$, and we have supposed $\ell' \geq 2 \gamma'+2$
and $E(r+d)\leq r_1$. Now, in summary, this yields
\begin{equation}
\lnorm{\oprod - \up \oprod}{\ell} = d^{-\ell} O(r) + O(r^{-k}) \big( (E(r+d))^{\gamma'} + (1+E)^{-\ell'/2} \big).
\end{equation}
For given $\beta > 0$, we now choose $r = d^{\ell+\beta+1}$, $E = r_1\, d^{-1/2}$, 
$\gamma' = (2k+2)(\ell+\beta+1)$, and $\ell' = 2 \gamma'+2$. 
With this choice, we obtain
\begin{equation}
\lnorm{\oprod - \up \oprod}{\ell} = o(d^{\beta}).
\end{equation}
We summarize our result as follows:

\begin{theorem} \label{opeUniformThm}
Let $\oprod$ be defined as in Eq.~\eqref{oprodDef}. 
%\todo{add more details? summarize conditions for theorem} 
For every $\beta > 0$, there exist $\gamma' > 0$ and $\ell>0$ such that,
with $\up$ being a uniform projector onto $\Phi_{\gamma'}$ as in Proposition~\ref{uniformProjProp},
\begin{equation*}
  d^{-\beta} \lnorm{\oprod - \up \oprod}{\ell} \to 0 \quad \text{as } d \to 0.
\end{equation*}
\end{theorem}

This constitutes a uniform OPE at all scales. We now transfer these estimates
to the limit theory. The OPE terms in the limit are supposed to be
$\pi_0 \up \oprod$, i.e. the limit of those at finite scales,
and they should approximate $\pi_0 \oprod$, the limit of the product.
First, we show that $\pi_0$ is in fact compatible with the product structure.
\begin{lemma} \label{productLimitLemm}
For any $f,f' \in \scal(\mkr)$ and any $d>0$, we have:
\begin{equation*}
    \pi_0 \oprod = \big(\alpha_0[f_d] (\pi_0\uphi) \big) \cdot \big(\alpha_0[f_d'] (\pi_0\uphi') \big).
\end{equation*}
\end{lemma}
\begin{proof}
We certainly know that an analogous relation holds between bounded operators:
\begin{equation} 
    \pi_0 \uB_r = \big(\alpha_0[f_d] (\pi_0 \uA_r) \big) \cdot \big(\alpha_0[f_d'] (\pi_0\uA_r') \big).
\end{equation}
Considering $\uB_r$ as an element of $\uPhi$, we obtain by Proposition~\ref{phiLimitProp} and Lemma~\ref{prodApproxLemm},
\begin{equation} \label{limitProdApprox}
  \lnorm{ \pi_0 (\uB_r - \oprod) }{2\ell+1} \leq  \lnorm{ \uB_r - \oprod }{\ell} \cdot \text{const} \to 0
\quad \text{as } r \to 0,
\end{equation}
where $\ell$ is large enough, and $d$ is kept fixed. Also, by Theorem~\ref{fieldApproxThm} and Eq.~\eqref{phiLimitSymm},
we know that for large $\ell$,
\begin{equation}
 \lnorm{ \alpha_0[f_d] \pi_0(\uA_r) - \alpha_0[f_d] \pi_0(\uphi) }{\ell} \to 0
\quad \text{as } r \to 0.
\end{equation}
The same holds for $\uA'$, $\uphi'$ in place of $\uA$, $\uphi$.
We can now use techniques as in the proof of Lemma~\ref{prodApproxLemm} to show that, for large $\ell$,
\begin{equation}
 \lnorm{ \pi_0(\uB_r) - \big(\alpha_0[f_d] \pi_0(\uphi)\big)\big( \alpha_0[f_d'] \pi_0(\uphi')  \big) }{\ell} \to 0
\quad \text{as } r \to 0.
\end{equation}
Combined with Eq.~\eqref{limitProdApprox}, this yields the proposed result. 
\cmpqed\end{proof}

Now applying Proposition~\ref{phiLimitProp} to the result of Theorem~\ref{opeUniformThm},
we can summarize our results as follows:

\begin{corollary} \label{opeLimitCor}
For every $\beta > 0$, there exists $\gamma'>0$, $\ell>0$, and a uniform projector $\up$ onto $\uPhi_{\gamma'}$ 
such that 
\begin{equation*}
   d^{-\beta} \lnorm{ \pi_0\oprod - \pi_0 \up \oprod }{\ell} \to 0
\quad \text{as }d \to 0.
\end{equation*}
Here we have
\begin{equation*}
    \pi_0 \oprod = \big(\alpha_0[f_d] (\pi_0\uphi) \big) \cdot \big(\alpha_0[f_d'] (\pi_0\uphi') \big).
\end{equation*}
Further, $\pi_0(\up \oprod) \in \PhiFHLim$ at any fixed $d$.
\end{corollary}

The last part follows by applying Corollary~\ref{limitFieldCor} to $\up\oprod$.

We may interpret these results as follows: The product of fields at fixed scales
converges to the corresponding product of fields in the limit theory;
and the OPE terms at fixed scales converge to OPE terms in the limit theory.
If $\mean$ is multiplicative, then we obtain finitely many independent OPE terms, 
in the sense that the multilinear map $(\uA,\uA',f,f') \mapsto \pi_0\oprod = 
\pi_0\up(\ualpha[f]\psi\st(\uA)\ualpha[f']\psi\st(\uA'))$ has a
finite-dimensional image if we keep the approximation map $\psi$ fixed. If $\mean$ is not multiplicative,
the OPE may be degenerate in the sense discussed in Sec.~\ref{renormSec}. 

In order to understand the role of the renormalization factors in the OPE, let us again translate
our results to the usual notation in physics. From Corollary~\ref{opeLimitCor}, the OPE terms
in the limit theory are given by $\pi_0 \up \oprod$. We write each $\up_\lambda$ in a basis:
$\up_\lambda = \sum_j \sigma_{j,\lambda}(\cdotarg) \phi_j$, where we can choose the fields $\phi_j$
independent of $\lambda$, but the functionals $\sigma_{j,\lambda}$ will generally depend on $\lambda$.
Writing the distributions as formal integration kernels (as usual in physics), this gives
\begin{equation}
  (\up \oprod)_\lambda = 
  \big( \up ((\ualpha_x \uphi)(\ualpha_{x'} \uphi')) \big)_\lambda
  = \sum_{j} \sigma_{j,\lambda}( (\ualpha_x\uphi)_\lambda (\ualpha_{x'}\uphi')_\lambda ) \phi_j.
\end{equation}
So the expressions $c_{j,\lambda}(x,x') := \sigma_{j,\lambda}( (\ualpha_x\uphi)_\lambda (\ualpha_{x'}\uphi')_\lambda )$ can be interpreted as OPE coefficients at scale $\lambda$.
For further comparison with perturbation theory, let us choose
$\uphi_\lambda^{(m)} = \sum_n Z_{n,\lambda}^{(m)} \phi_n$ (see
Eq.~\eqref{renormFactor}) such that at scale 1, we have $Z_{n,1}^{(m)}=\delta_{mn}$. Let us assume that the matrix $Z_{n,\lambda}^{(m)}$ is invertible at each fixed $\lambda$.
The product expansion for the product $\oprod^{(m,m')}$ with $\uphi=\uphi^{(m)}$ and $\uphi'=\uphi^{(m')}$ 
then reads
\begin{equation} \label{zCoeffSum}
  (\up \oprod^{(m,m')})_\lambda 
  = \sum_{j} 
  \underbrace{ \sum_{k,n,n'} Z_{n,\lambda}^{(m)} Z_{n',\lambda}^{(m')} 
  (Z^{-1})_{j,\lambda}^{(k)} \sigma_{k,\lambda}( \phi_n (\lambda x)  \phi_{n'} (\lambda x') ) }_{=: c_{j,\lambda}^{(m,m')}(x,x')}  \uphi_{\lambda}^{(j)}.
\end{equation}
\emph{If we can assume here} that the functionals $\sigma_{k,\lambda}$ can be chosen independent
of $\lambda$, then this gives the formula well known from perturbation theory (cf.~Eq.~(98) in \cite{Hol:ope_curved}):
\begin{equation} \label{hollandsComparison}
     c_{j,\lambda}^{(m,m')} (x,x') = \sum_{k,n,n'} Z_{n,\lambda}^{(m)} Z_{n',\lambda}^{(m')} (Z^{-1})_{j,\lambda}^{(k)} \, c_{k,1}^{(n,n')}( \lambda x, \lambda x' ).
\end{equation}
Whether the assumption of $\lambda$-independent functionals $\sigma_j$ is justified remains open, however.
In fact, if the basis functionals $\sigma_{j,\lambda}$ are chosen energy-bounded, which is always possible at finite scales,
one would rather expect that this energy bound needs to be properly rescaled as $\lambda \to 0$. 
However, in the context of perturbation theory, it may be justified to assume that
$\sigma_{j,\lambda}$ is independent of $\lambda$ up to terms of higher order.

The symmetry group $\ugcal$ of the scaling algebra acts on the uniform operator product expansions
in a natural way. Namely, Let $g=(\mu,0,\Lambda)$ be a dilation and/or Lorentz transform, but with
no translation part. By Theorem~\ref{opeUniformThm}, the OPE is given by 
\begin{equation} \label{oprodWoSymm}
  d^{-\beta} \lnorm{ \oprod - \up \oprod}{\ell} \to 0 \quad \text{as } d \to 0.
\end{equation}
Now note that,
if $\up$ is a uniform projector as described in Proposition~\ref{uniformProjProp},
then $\ualpha_g^{-1} \up \ualpha_g$ is also a projector with the properties 
given in that lemma. So we also obtain
\begin{equation}\label{oprodWithSymm}
 d^{-\beta} \lnorm{  \oprod  - \ualpha_g^{-1} \up \ualpha_g \oprod}{\ell} \to 0
\quad \text{as } d \to 0.
\end{equation}
Combining Eqs.~\eqref{oprodWoSymm} and \eqref{oprodWithSymm}, and observing that we
can apply $\ualpha_g$ within the norm without changing the limit, we obtain
\begin{equation} \label{opeSymmStatement}
 d^{-\beta} \lnorm{  \ualpha_g \up \oprod - \up \ualpha_g \oprod }{\ell} \to 0
\quad \text{as } d \to 0.
\end{equation}
In other words, the OPE terms $\up \ualpha_g \oprod$ for the transformed product
$\ualpha_g \oprod$ are the same as the $\ualpha_g$-transformed OPE terms $\up\oprod$ for $\oprod$,
up to terms of higher order in $d$. The same holds then in the limit theory,
by application of $\pi_0$ to all terms. As above, we can use a basis representation of $\up$ 
and $\uphi, \uphi'$ in order to express the relation in Eq.~\eqref{opeSymmStatement} 
in terms of renormalization factors. It should be noted that, in a perturbative context,
symmetry properties of the OPE are one of its key features that is exploited for
applications; see e.g. \cite{BDLW:sum_rules}.

The situation is simpler if we consider only products evaluated in the vacuum state,
i.e. if we compute the renormalization limits of Wightman functions. We shall only
consider the case of a two-point function here, with $\uphi'=\uphi\st$; other $n$-point
functions can be handled in a similar way. 
Evaluating the results of Lemma~\ref{productLimitLemm} in the vacuum (for $d=1$), 
we have for $\phi_0 := \pi_0 \uphi$,
\begin{equation}
 \mtxe{\Omega_0}{ (\alpha_{0}[f] \phi_0) (\alpha_{0}[f'] \phi_0\st )}{\Omega_0}
=  \mean ( \omega( (\ualpha_f \uphi)_\lambda (\ualpha_{f'} \uphi)_\lambda\st ))
\end{equation}
Defining the usual two-point functions $W_{jk}(x,x') = \mtxe{\Omega}{\phi_j(x)\phi_k(x')}{\Omega}$ in the original
theory, and $W_0(x,x') = \mtxe{\Omega_0}{\phi_0(x)\phi_0\st(x')}{\Omega_0}$ the limit theory, 
we obtain:
\begin{equation}
 W_0(x,x') =  \mean \big( \sum_{j,k} Z_{j,\lambda} \bar Z_{k,\lambda} W_{jk} (\lambda x, \lambda x') \big).
\end{equation}
So the Wightman functions ``converge'' (in the sense of means) to their expected limits.
To illustrate the consequences, let us consider an invariant mean $\mean$,
and let us assume the following simplified situation:
\begin{enumerate}
\localitemlabels
\item We only deal with \emph{one} renormalization factor, i.e. $\uphi_\lambda = Z_\lambda\phi$
with a fixed $\phi \in \cinftyss$. (We then have only one Wightman function $W_{11}=W$ at 
finite scales.)
\item The factor $|Z_\lambda|$ is strictly positive and monotonously decreasing as $\lambda \to 0$.
\item $|Z_\lambda|^2 W(\lambda x, \lambda x')$ \emph{converges} to $W_0(x,x')$ in
      the topology of $\scal'$ as $\lambda \to 0$ (not only in the sense of means).
\item $W_0$ is not the zero distribution.
\end{enumerate}
Then, for any $\mu \geq 1$, the function $\lambda \mapsto Z_{\lambda/\mu}/Z_\lambda$ is bounded;
we set $h(\mu) := \mean(|Z_{\lambda/\mu}/Z_\lambda|^2)$. We know that, in the
sense of distributions,
\begin{equation}
  W_0(\mu x, \mu x') = \mean( |Z_\lambda|^2 W(\mu\lambda x, \mu\lambda x') )
 = \mean( |Z_{\lambda/\mu}|^2 W(\lambda x, \lambda x') )
 = h(\mu) W_0(x,x').
\end{equation}
Here we haved used the invariance of the mean, and the fact that the mean ``factorizes'' 
since $|Z_\lambda|^2 W(\lambda x, \lambda x')$ is convergent by assumption.
It is clear from the above that $h(1) = 1$ and $h(\mu\mu') = h(\mu)h(\mu')$. Also, 
$h$ is continuous since $\mu \mapsto W_0(\mu x, \mu x')$ is continuous in $\scal'$,
and since $W_0 \neq 0$. As is well known, this implies $h(\mu)= \mu^{-a}$ for some $a \geq 0$.
This reproduces a result from \cite{FreHaa:covariant_scaling} in our context.
Note that it is not implied that $|Z_\lambda|^2 = \lambda^{a}$; rather, $Z_\lambda$ might also differ from
this by a slowly varying factor, such as $|Z_\lambda|^2 = \lambda^{a} (\log \lambda)^b$.

We have confined our attention here to the case of a product of two fields, in the sense of 
distributions. Many generalizations of this setting are certainly within reach. First, the analogue
of Corollary~\ref{opeLimitCor} should hold for products of an arbitrary finite number of fields, 
and for their linear combinations. 
One can also allow more general short distance limits than a simple scaling of the test functions $f_d$,
at the price of increased technical effort. Moreover, like shown for the theory at fixed scales
in \cite{Bos:product_expansions}, it should be possible to obtain more detailed results on the OPE
at spacelike distances, where the OPE coefficients exist as analytic functions rather than only as distributions.
This would be particularly interesting for obtaining estimates on the two-point function and its limit, 
which might lead to a criterion for asymptotic freedom, since massless free theories can be characterized
by estimates on their two-point function \cite{Poh:zero_mass_fields,DAn:dilations,Bau:two_point_functions}. 
These extensions go beyond the scope
of the current paper however; we hope to return to these questions elsewhere.

\section{Conclusions}  \label{conclusionsSec}
The renormalization group methods in quantum field theory have found their main applications
in the study of short distance properties of non-abelian gauge theories,
such as quantum chromodynamics (QCD), which are expected to exhibit interesting features like
confinement and asymptotic freedom. Since no rigorously constructed 
version of QCD is available to date, it is not clear if the result presented here 
are directly applicable to this case; but some heuristic comments are anyway in order.

According to our results in Theorem~\ref{limitApscThm}, ``no new fields appear
in the scaling limit''. In view of the common expectations about the confining
dynamics of QCD, it may seem that this would imply that our phase space
conditions are not general enough to encompass such a theory. 
However, this conclusion is not justified, since we are restricting attention here
to \emph{observable} fields only, and do not directly deal with charged fields. This is also 
not necessary, since the gauge group and the 
field algebras\footnote{%
In order to avoid confusion in terminology, let us note that the field algebras
$\ffk(\ocal)$ in the sense of Doplicher-Roberts are supposed to contain
\emph{charge-carrying} objects, but still bounded operators associated with
an open region $\ocal$, not (unbounded) \emph{pointlike localized} fields.}
of unobservable objects can be constructed from the
algebras of observables by means of charge analysis~\cite{DopRob:field_algebra}.
More precisely, the following diagram of theories holds:
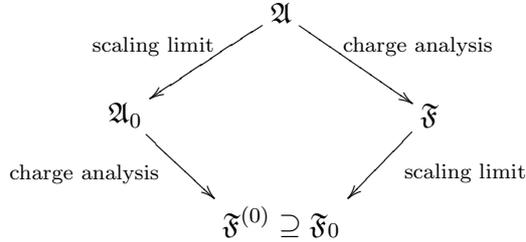
\begin{equation*}
%\spreaddiagramrows{-7pt}
%\spreaddiagramcolumns{-21pt}
\xymatrix{
      & \afk \ar[dl]_{\text{ scaling limit }} \ar[dr]^{ \text{ charge analysis } } \\
          \afk_0 \ar[dr]!UL_{\text{ charge analysis }} & 
           &\ffk \ar[dl]!UR^{\text{ scaling limit }}\\
    & \ffk^{(0)} \supseteq \ffk_0
    }
\end{equation*}
Here $\ffk$ is the field algebra at finite scales, in the sense of Doplicher-Roberts,
$\ffk_0$ is its scaling limit, and $\ffk^{(0)}$ is the field algebra constructed 
from the scaling limit of observables, $\afk_0$. 
The inclusion in the bottom line of the diagram is a
strict one in the case of confinement~\cite{Buc:quarks}.\footnote{A general discussion of the relations between
the superselection structures of $\afk$ and $\afk_0$ and the corresponding Doplicher-Roberts
field nets, leading to an intrisic notion of charge confinement, has been performed 
in~\cite{Buc:quarks,DMV:scaling_charges,DAnMor:supersel_models}, to which we refer the interested reader
for further details.}
In the case of non-abelian gauge theories, it is 
generally expected that in covariant gauges, charged fields of the theory
at finite scales, such as
the field strength tensor $F_{\mu\nu}^a(x)$ and the quark fields $Q^a(x)$ (here $a$ denotes the SU(3)
color index),
are non-local when restricted to the ``physical'' Hilbert
space of the states satisfying the gauge condition (see e.g.~\cite{Str:generalQFT}).
In particular these fields are not observable, even if some of their functions,
such as e.g. $F_{\mu\nu}^a(x) F_a^{\mu\nu}(x)$ in the sense of some normal product, 
may be. On the other hand, assuming
that the traditional scenario of asymptotic freedom holds beyond perturbation
theory, in the scaling limit we expect that the corresponding fields $F_{0,\mu\nu}^a(x)$, $Q_0^a(x)$
are free local massless fields. Still they are not observable, as they transform
nontrivially under color SU(3), which is now a true (global) symmetry in the ultraviolet
limit. Charged fields such as $F_{0,\mu\nu}^a(x)$ and
$Q_0^a(x)$ can then be regarded as being associated with $\ffk^{(0)}$, but not with $\ffk_0$,
as they should not be the limit of pointlike fields associated to $\ffk$. It should also be kept in mind
that it is in general necessary,  in order to perform the superselection analysis in the scaling
limit, to pass from $\afk_0$ to its dual net $\afk_0^d$, and it is thus possible that new fields
appear there. In view of these facts, our result about the non-increase of the
number of observable fields when passing to the scaling limit seems to be in
line with the general picture of confinement in QCD.

The approach to the renormalization of quantum fields that we presented here is more general
than the traditional one, including also cases where the scaling limit is not unique.
Also, we have shown that the very existence of renormalization factors, which is an ansatz in the traditional approach, is actually
a consequence of the general properties of quantum field theory.
In our context it is always possible, 
independent of the model under consideration,
to form finite linear combinations of the fields associated to the finite scale theory, with suitable, 
scale dependent coefficients, in such a way that the resulting field has a well-defined
limit as a field associated to the scaling limit theory. It should be stressed that in our approach
no requirement is made about the convergence of $n$-point functions at small scales, so that
our results are applicable also to the situations in which no proper fixed points of the
renormalization group exist. Such theories are those with a ``degenerate scaling limit''
according to the classification of~\cite{BucVer:scaling_algebras}. The short
distance behavior of these theories is described by a whole family of scaling limits, distinguished
by the choice of a mean along which the limit is performed. 

We also have considered
the behavior of operator product expansions under scaling. We have shown that it is possible
to obtain a uniform expansion at all scales, which converges to the 
expansion of the product of the limit fields, and whose coefficients satisfy, at least in 
special cases, the scaling law which is customary in perturbation theory, cf.\ Eq.~\eqref{hollandsComparison}.

Our analysis uses as a basic input the phase space condition \ref{phase2} stated in Sec.~\ref{fieldSec}, 
allowing the identification of the pointlike fields associated to the given net of local algebras. 
The validity of such condition has been verified in models with a finite number
of free fields, massive or massless, in $s \geq 3$ spatial dimensions~\cite{Bos:operatorprodukte}.  
In view of these facts, it seems reasonable to
expect that this criterion is verified also in more general field theoretical models, possibly
interacting, in particular in asymptotically free theories, since their short-distance behavior should
not differ significantly from that in free models.

In order to find a counterpart of the action of the Gell-Mann and Low renormalization 
group in our framework,
we had to generalize the notion of scaling limit given in~\cite{BucVer:scaling_algebras},
introducing a class of dilation invariant but not pure scaling limit states. The study of the
structure of the scaling limit theory corresponding to such states is in progress~\cite{BDM:scaling2}.
As mentioned in Section~\ref{scalingSec}, it is possible to show that, for the theory
of a massive free scalar field, such scaling limit theory is a tensor product 
of the algebras of the massless free field with a model-independent abelian factor, which
corresponds to the restriction of the scaling limit representation to the center of the scaling algebra. In general, scaling limits with the described tensor product structure
fall into the class of theories with unique vacuum structure, in the terminology of \cite{BucVer:scaling_algebras}.
The precise conditions under which such a tensor product structure occurs
are currently under investigation.

However, there are certainly other models which do not exhibit a simple tensor product
structure in the limit. In this context, it seems interesting to investigate
the model proposed in~\cite{BucVer:scaling_examples}, i.e.\ an infinite tensor product of free fields with masses $2^n m$, $n \in \zbb$.
Although this model obviously violates our phase space conditions, so that the analysis of the 
scaling behavior of its field content is out of reach with the methods employed here, its scaling
limit can nevertheless be considered from the algebraic point of view, and it could give
an interesting example of a dilation covariant theory 
which contains components of massive free fields. Finally, also in view of the 
discussion above about possible applications of the present analysis to physically interesting
models, it would be worthwhile to extend the results presented here to treat the renormalization
of charge carrying pointlike fields, associated with the Doplicher-Roberts field net.

\appendix

\section{The normal part of a functional} \label{normalPartApp}

For our investigation, we need the concept of the \emph{normal part} of
a functional $\rho$ in the dual of some $C\st$ algebra $\afk$. 
That is, we want to extract from $\rho$ that part which is in the folium
of some given state $\omega$. The techniques used in this context
can be found in \cite[Ch.~10.1]{KadRin:algebras2}; 
we will however repeat some part of the construction here, since
we need some properties specific to our setup.

\begin{theorem}\label{normalpartThm}
Let $\afk$ be a $C\st$ algebra, $\omega$ a state on $\afk$, and 
$\pi_\omega$ the associated GNS representation. Denote by $\Sigma_\omega$
the space of ultraweakly continuous functionals on $\pi_\omega(\afk)''$. 
There exists a linear map $\normalpart_{\omega}: \afk\st \to \Sigma_\omega$
with the following properties.\footnote{%
With $\pi$ being a representation of a $C\st$ algebra, we denote
by $\pi\st$ the ``pullback'' action on the dual spaces:
$\pi\st: \pi(\afk)\st \to \afk\st$, $\rho \mapsto \rho \circ \pi$.}
\begin{enumerate}
\localitemlabels
\item \label{normalpartIdItem}

$\normalpart_\omega \circ \pi_\omega\st \restrict \Sigma_\omega = \mathrm{id} \restrict \Sigma_\omega$.

\item \label{normalpartPositiveItem}

If $\rho \in \afk\st$, $\rho \geq 0$, then 
$\normalpart_\omega(\rho) \geq 0$ and $\rho \geq \pi_\omega\st \normalpart_\omega(\rho)$.

\item \label{normalpartNormItem}
$\|\normalpart_\omega\| = 1$.

\end{enumerate}
$\normalpart_\omega$ is uniquely determined by properties~\ref{normalpartIdItem}
and \ref{normalpartPositiveItem}.
\end{theorem}

We will call $\normalpart_{\omega}(\rho)$ the \emph{normal part of $\rho$ with respect to $\omega$.}
It depends on the algebra $\afk$ and is usually not compatible with the restriction to subalgebras.
Therefore, we will label the normal part also as $\normalpart[\afk,\omega](\rho)$, where the reference
to the base algebra $\afk$ is particularly important. 

\begin{proof}
We first show uniqueness. Let $\normalpart_\omega$, $\tilde \normalpart_\omega$ be two maps which fulfill
\ref{normalpartIdItem} and \ref{normalpartPositiveItem}. Set $Q = \pi_\omega\st \normalpart_\omega$, 
$\tilde Q = \pi_\omega\st \tilde\normalpart_\omega$. From \ref{normalpartIdItem}, one easily sees that
$\tilde Q Q = Q$. Now let $\rho \in \afk\st$, $\rho \geq 0$. Due to \ref{normalpartPositiveItem},
one has $\rho \geq Q \rho$, thus $\rho - Q \rho \geq 0$. Applying $\tilde Q$ to this positive
functional, and observing that $\tilde Q$ preserves positivity due to \ref{normalpartPositiveItem},
we have
\begin{equation}
   \tilde Q (\rho - Q \rho) \geq 0
\csq
  \tilde Q \rho - \tilde Q Q \rho \geq 0
\csq
  \tilde Q \rho \geq Q \rho.
\end{equation}
By symmetry, however, we likewise obtain $Q \rho \geq \tilde Q \rho$, thus $Q \rho = \tilde Q \rho$.
Since $\pi_\omega\st$ is clearly injective, it follows that $\normalpart_\omega \rho = \tilde\normalpart_\omega \rho$
for all positive $\rho$. By taking linear combinations, the same holds for all $\rho$.

Now for existence: Let $\pi\universal: \afk \to \bcal(\hcal\universal)$ be the universal representation of $\afk$.
For each $\rho \in \afk\st$, there is a unique ultraweakly continuous $\rho\universal \in (\pi\universal(\afk)'')\st$
such that $\rho\universal \circ \pi\universal = \rho$ \cite[p.~721]{KadRin:algebras2}.
Here the map $\rho \mapsto \rho\universal$ is linear, isometric, and preserves positivity.
Now according to \cite[Theorem~10.1.12]{KadRin:algebras2}, there exists an orthogonal projection
$P$ in the center of $\pi\universal(\afk)''$ and an ultraweakly bi-continuous isomorphism 
$\alpha: P \pi\universal (\afk)'' \to \pi_\omega(\afk)''$ such that
$\pi_\omega = \alpha \circ \mu_P \circ \pi\universal$, where $\mu_P: \pi\universal(\afk)''\to\pi\universal(\afk)''$
is the multiplication with $P$. We set
\begin{equation}
   \normalpart_\omega(\rho) := \rho\universal \circ \alpha^{-1} .
\end{equation}
This expression is ultraweakly continuous on  $\pi (\afk)''$; hence $\normalpart_\omega(\rho) \in \Sigma_\omega$.
Note that $\alpha^{-1}$ is norm preserving as an isomorphism, 
and $\|\rho_u\|= \|\rho\|$; thus $\|\normalpart_\omega\| \leq 1$. In fact, evaluating 
$\normalpart_\omega$ on $\omega$
yields $\|\normalpart_\omega\| = 1$, so we obtain \ref{normalpartNormItem}.

For $\sigma \in \Sigma_\omega$, and $\rho := \sigma \circ \pi_\omega$,
one has
\begin{equation}
  \rho\universal \circ \pi\universal = \sigma \circ \pi_\omega = \sigma \circ \alpha \circ \mu_P \circ \pi\universal;
\end{equation}
thus $\rho\universal = \sigma \circ \alpha \circ \mu_P$. It follows that
\begin{equation}
  \normalpart_\omega (\sigma \circ \pi_\omega) = \rho\universal \circ \alpha^{-1}
  = \sigma \circ \alpha \circ \mu_P \circ \alpha^{-1} = \sigma,
\end{equation}
for $\mu_P$ acts as identity on the image of $\alpha^{-1}$. This proves \ref{normalpartIdItem}.

Now let $\rho \in \afk\st$, $\rho \geq 0$, thus also $\rho\universal \geq 0$. Since
the isomorphism $\alpha^{-1}$ preserves positivity, one has $\normalpart_{\omega}(\rho) \geq 0$.
Further, note that
\begin{equation}
  \pi_\omega\st \normalpart_\omega(\rho)
 = \rho\universal \circ \alpha^{-1} \circ \pi_\omega 
 = \rho\universal \circ \mu_P \circ \pi\universal .
\end{equation}
It follows that
\begin{equation}
  \rho - \pi_\omega\st \normalpart_\omega(\rho)
 = \rho\universal \circ \pi\universal - \rho\universal \circ \mu_P\circ \pi\universal 
=  \rho\universal \circ (1-\mu_P) \circ \pi\universal.
\end{equation}
Using $P\st=P$, it is easy to see that $(1-\mu_P)$ preserves positivity, and so does $\pi\universal$
as a representation. Thus $\rho - \pi_\omega\st \normalpart_\omega(\rho) \geq 0$.
This proves \ref{normalpartPositiveItem}.
\end{proof}

\section*{Acknowledgements}
The authors would like to thank K.~Fredenhagen, G.~Morchio and F.~Strocchi for discussions on the subject,
and D. Buchholz for information on his previous work on the topic. 

A substantial part of this work was done during stays of the authors at the Erwin Schrödinger Institute,
Vienna. H.B. wishes to thank the Universities of Rome ``La Sapienza'' and ``Tor Vergata'' for their hospitality.
He would also like to thank the University of Florida for an invitation.

%\todo{expand acknowledgements}

\bibliographystyle{alpha}
\bibliography{qft}

\end{document}